\documentclass[useAMS,usenatbib,usegraphicx]{mn2e}

\title[Vibronic bands of C$_3$ towards HD~169454]
  {Detection of vibronic bands of C$_3$ in a translucent cloud towards HD 169454
}
\author[M.~R.~Schmidt et al.]
{M.~R.~Schmidt,$^1$\thanks{email: schmidt@ncac.torun.pl}
  J.~Kre{\l}owski,$^2$
  G.~A.~Galazutdinov,$^{3,4}$\thanks{email: runizag@gmail.com}
\newauthor
 D. Zhao,$^5$
 M.~A.~Haddad,$^6$
 W.~Ubachs,$^6$
 H.~Linnartz$^5$\\
  $^1$Department of Astrophysics, N. Copernicus Astronomical Center,
  ul. Rabia{\'n}ska 8, 87-100 Toru{\'n}, Poland \\
  $^2$Centre for Astronomy, Nicolaus Copernicus University, Gagarina 11, 87-100 Toru{\'n}, Poland\\
  $^3$Instituto de Astronomia, Universidad Catolica del Norte, Av. Angamos 0610,
  Antofagasta, Chile\\
  $^4$Pulkovo Observatory, Pulkovskoe Shosse 65, Saint-Petersburg 196140, Russia\\
  $^5$Raymond and Beverly Sackler Laboratory for Astrophysics, Leiden Observatory, Leiden University,
  PO Box 9513,\\ 2300 RA Leiden, The Netherlands\\
  $^6$Department of Physics and Astronomy, LaserLaB, VU University,
  De Boelelaan 1081, 1081 HV Amsterdam, The Netherlands\\
}
\date{Date: 2013 Xxxxx XX}

\pagerange{\pageref{firstpage}--\pageref{lastpage}} \pubyear{2012}

\def\LaTeX{L\kern-.36em\raise.3ex\hbox{a}\kern-.15em
    T\kern-.1667em\lower.7ex\hbox{E}\kern-.125emX}

\begin{document}

\label{firstpage}

\maketitle

\begin{abstract}

We report the detection of eight vibronic bands of C$_3$, seven of which
have been hitherto unobserved in astrophysical objects, in the translucent
cloud towards HD~169454.  Four of these bands are also found towards two
additional objects: HD~73882 and HD~154368. Very high signal-to-noise ratio
($\sim$1000 and higher) and high resolving power ($R=80,000$) UVES-VLT
spectra (Paranal, Chile)
% {\bf also}
allow for detecting novel spectral features of C$_3$,
even revealing weak perturbed features in the strongest bands.
The work presented here provides the most complete spectroscopic survey of the so far largest
carbon chain detected in translucent interstellar clouds. High-quality laboratory
spectra of C$_3$ are measured using cavity ring-down absorption spectroscopy
in a supersonically expanding hydrocarbon plasma, to support the analysis
of the identified bands towards HD~169454.
A column density of N(C$_3$) = $(6.6 \pm 0.2) \times 10^{12}$ cm$^{-2}$ is inferred
and the excitation of the molecule exhibits two temperature components;
$T_{exc}= 22 \pm 1$ K for the low-$J$ states and $T_{exc}= 187 \pm 25$ K
for the high-$J$ tail. The rotational excitation of C$_3$ is reasonably well
explained by models involving a mechanism including inelastic collisions,
formation and destruction of the molecule, and radiative pumping in the far-infrared.
% These models yield a slightly lower collisional temperature below $18 \pm 1$ K,
% which is attributed to C$_3$ being formed towards the central region of the diffuse cloud.
% These models {\bf suggest} a slightly lower {\bf gas kinetic} temperature below $18 \pm 1$ K,
% which {\bf may be} attributed to C$_3$ being formed towards the central region of the diffuse cloud.
These models yield gas kinetic temperatures comparable to those found for $T_{exc}$.
The assignment of spectral features in the UV-blue range 3793-4054 \AA\
may be of relevance for future studies aiming at unravelling spectra
to identify interstellar molecules associated with the diffuse
interstellar bands (DIBs).

\end{abstract}


\section{Introduction}

Currently, some 180 different molecules have been detected in
dense inter- and circumstellar clouds,
largely in radio- and submilimeter surveys,
with this number growing with several new species per year.
However, only about ten simple molecules are observed in the visible
part of the electromagnetic spectrum as absorption features originating
in translucent clouds, transparent for optical wavelengths.
Among them are homonuclear species, such as H$_2$, C$_2$ and C$_3$,
which are not accessible to radio observations.
Bare carbon chains do not exhibit pure rotational transitions,
because of the lack of a permanent dipole moment,
and thus only their electronic or vibrational spectral features can be observed.
The latter cover the spectral range from the vacuum UV until the far infrared.
Determination of the abundances of simple carbon molecules in interstellar clouds
is important, as they are considered building blocks for many already known
interstellar molecules with a carbon skeleton.

After the $19^{th}$ century discovery of the 4052 \AA\ band in the spectrum
of comet Tebbutt \citep{Huggins1881} and the assignment of this blue absorption
feature to the C$_3$ molecule \citep{Douglas1951}, the triatomic
carbon chain radical was detected  in the circumstellar shell of the star IRC+10216 \citep{Hinkle1988}
and subsequently, albeit tentatively, in the interstellar medium \citep{Haffner1995}.
\citet{Cernicharo2000} detected nine lines of the $\nu_2$ bending mode towards Sgr B2 and
IRC+10216, observed in the laboratory by~\citet{Giesen2001}.
This spectral range is, however, not applicable to observations of translucent clouds.
High resolution observations
using the Herschel/HIFI instrument detected transitions of C$_{3}$
originating in the warm envelopes of massive star forming
regions~\citep{Mookerjea2010, Mookerjea2012};
in these environments high densities of $10^5-10^6$ cm$^{-3}$ prevail,
whereas in diffuse clouds densities are limited to $10^3$ cm$^{-3}$.
The presence of C$_3$ in the diffuse interstellar medium was proven by
\citet{Maier2001},
based on the detection of the \~{A}$^{1}\Pi_{u}$--\~{X}$^{1}\Sigma_{g}^{+}$ 000--000 band
close to 4052~\AA\ towards three reddened stars.
Up to now the highest resolution study of C$_3$ in such environments
was reported by~\citet{Galazutdinov2002b}, although
restricted to a few objects only. These observations provided the only reliable
estimates so far of the abundance of this molecule, yielding values an order of magnitude below that of C$_2$.
Other observations of C$_3$ in translucent clouds~\citep{Roueff2002, Adamkovics2003, Oka2003}
suffered from lower signal-to-noise ratios, giving rise to large uncertainties in the deduced column densities.

The existing data for linear carbon molecules longer than C$_3$,
like C$_4$ \citep{Linnartz2000}, or C$_5$ \citep{Motylewski1999b},
fail to provide firm evidence for their existence in the interstellar medium
\citep{Galazutdinov2002a, Maier2002, Maier2004}.
For a systematic investigation of the conditions under which carbon-based molecules are produced
in the interstellar medium,
a larger class of targets exhibiting a variety of optical properties
and physical conditions in the intervening clouds has to be observed.
The targeted objects should be selected for translucent clouds with the
carbon-bearing molecules producing a single Doppler-velocity component,
thus allowing for an unambiguous analysis of the spectrum,
and resulting in accurate column densities.

The identification of the carriers of diffuse
interstellar bands (DIBs) remains, since their discovery by \citet{Heger1922},
one of the persistently unresolved problems in spectroscopy.
The current list of unidentified interstellar
absorption features contains more than 400 entries
\citep{Hobbs2008}. The presence of substructures inside DIB
profiles, discovered by \citet{Sarre1995} and by
\citet{Kerr1998}, supports the hypothesis of their molecular
origin. The established relation between profile widths of DIBs and
rotational temperatures of linear carbon molecules  \citep{Kazmierczak2010}
makes the latter interesting targets for observations.
Both C$_2$ and C$_3$ may show different
rotational temperatures along different lines of sight,
as shown by \citet{Adamkovics2003}.
This is associated with the fact that their rotational transitions are forbidden
and thus cooling of their internal degrees of freedom is inefficient.
For this reason, an accurate determination of rotational excitation
temperatures of short carbon chains may help to shed light on the origin of
the mysterious carriers of the DIBs \citep{Kazmierczak2010b}.

Observationally, the spectral features originating from either C$_2$ or
C$_3$ typically turn out to be rather shallow and thus high
S/N ratios and high spectral resolution are required to
establish accurate values for the excitation temperature.
The aim of the present investigation is to use the superior capabilities of UVES-VLT to
obtain high quality spectra of C$_3$.
All previous studies were based solely on the strongest 000--000 band of the
\~{A}$^{1}\Pi_{u}$--\~{X}$^{1}\Sigma_{g}^{+}$ electronic system.
Here additional vibronic bands in the
\~{A}$^{1}\Pi_{u}$--\~{X}$^{1}\Sigma_{g}^{+}$ electronic system of C$_3$
are identified along sight lines towards objects {HD~169454}, {HD~73882}, and {HD~154368}.
A detailed analysis of eight vibronic bands detected towards the object {HD~169454} is presented.
The astronomical observations are supported by a high-quality laboratory investigation,
using cavity ring-down laser spectroscopy, producing fully rotationally resolved C$_3$ spectra
of the vibronic bands in the \~{A}--\~{X} system.
The combined information of laboratory and observed spectra is used to deduce
column densities and a rotational temperature of C$_3$ in {HD~169454}
The results are interpreted in terms of excitation models for C$_3$
\citep{Roueff2002} and a chemical model of a translucent cloud towards HD~169454.

\section{Methods}

\subsection{Observations}

The observational material, of which a target list is presented in
Table~\ref{Table1}, was obtained using the UVES spectrograph mounted on the
ESO Very Large Telescope at Paranal (Chile) with resolution $R = 80,000$ in
the blue arm (3020 -- 4980~\AA) occupying the C$_3$ bands of interest,
as well as CH and CH$^+$.  The data set comprises spectra acquired during
our observation run of March 2009 (program 082.C-0566(A)) and data from the
ESO Archive programs 71.C-0367(A) and 076.C-0431(B).  The spectra, averaged
over 10 - 50 exposures are of exceptionally high S/N ratio with values
between 1900-2800.

\begin{table}
\caption{A list of observed targets in which C$_3$ was detected at 4052 \AA\
with specific physical characteristics and observed signal-to-noise ratio (S/N).}
% \label{targetslist}
\label{Table1}
\begin{tabular}{llllcc}
 Star        &  SpL     &   V    &  B-V     &  E(B-V)    &     S/N      \\
\hline
HD~73882    & O8V      &   7.21  & +0.40   &  0.67      &   $\sim$1900       \\
HD~154368   & O9.5Iab  &   6.14  & +0.50   &  0.73      &   $\sim$2200        \\
HD~169454   & B1Ia     &   6.62  & +0.90   &  1.11      &   $\sim$2800        \\
\hline
\end{tabular}
\end{table}

\begin{figure*}
\centering
\includegraphics[width=140mm, angle=270]{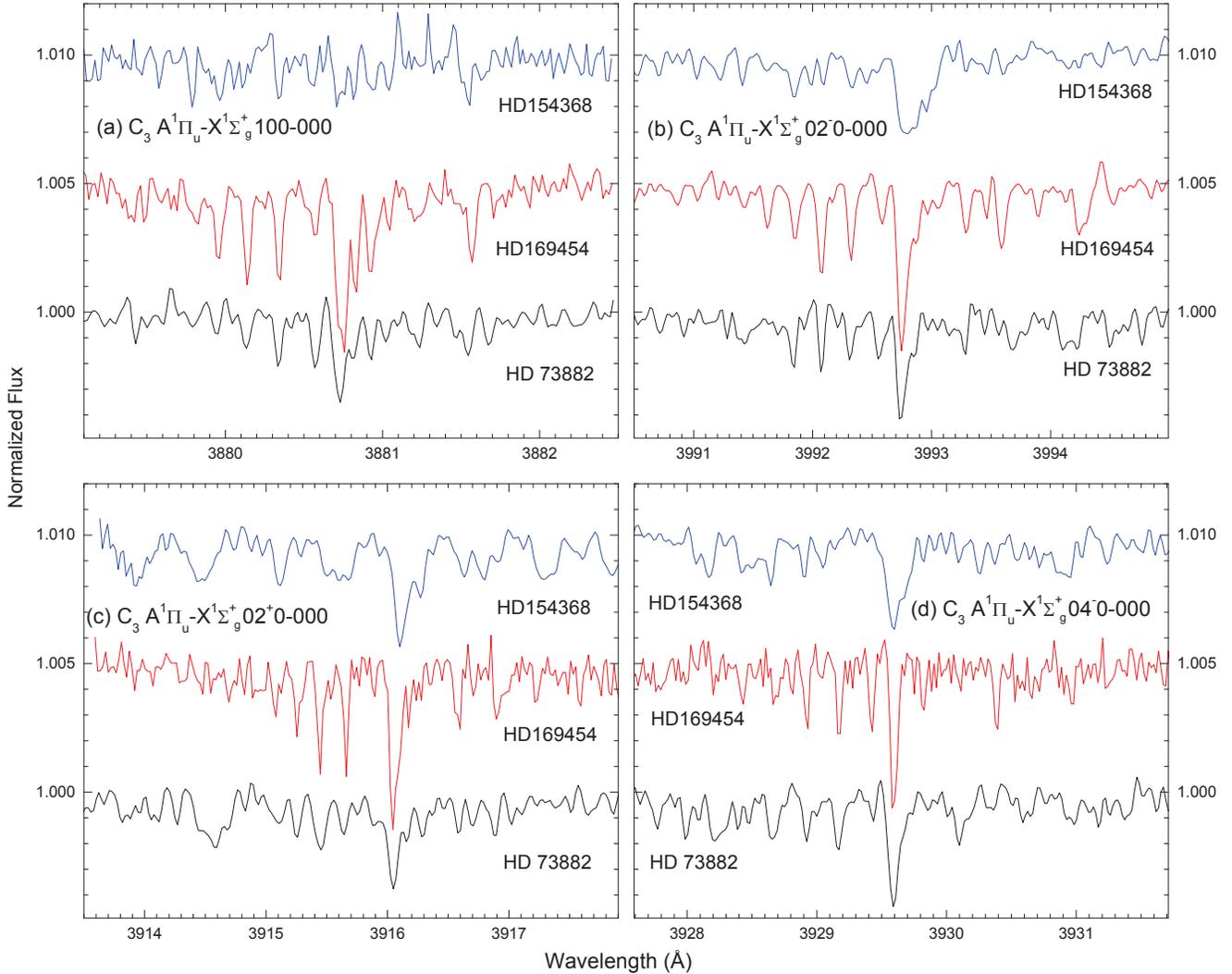}
\caption{
Four vibronic bands in the \~{A}$^{1}\Pi_{u}$--\~{X}$^{1}\Sigma_{g}^{+}$
system of C$_3$ observed in {HD~154368}, {HD~169454}, and {HD~73882}, .
}
\label{fig4bands}
\end{figure*}

All spectra were reduced with the standard IRAF packages, as well as our own
DECH code \citep{Galazutdinov1992}, providing the standard procedures of
image and spectra processing.  Using different computer codes for data
analysis reduces the inaccuracies connected to the slightly different ways
of dark subtraction, flat fielding, or excision of cosmic ray hits.

\begin{figure*}
\centering
\includegraphics[width=\textwidth]{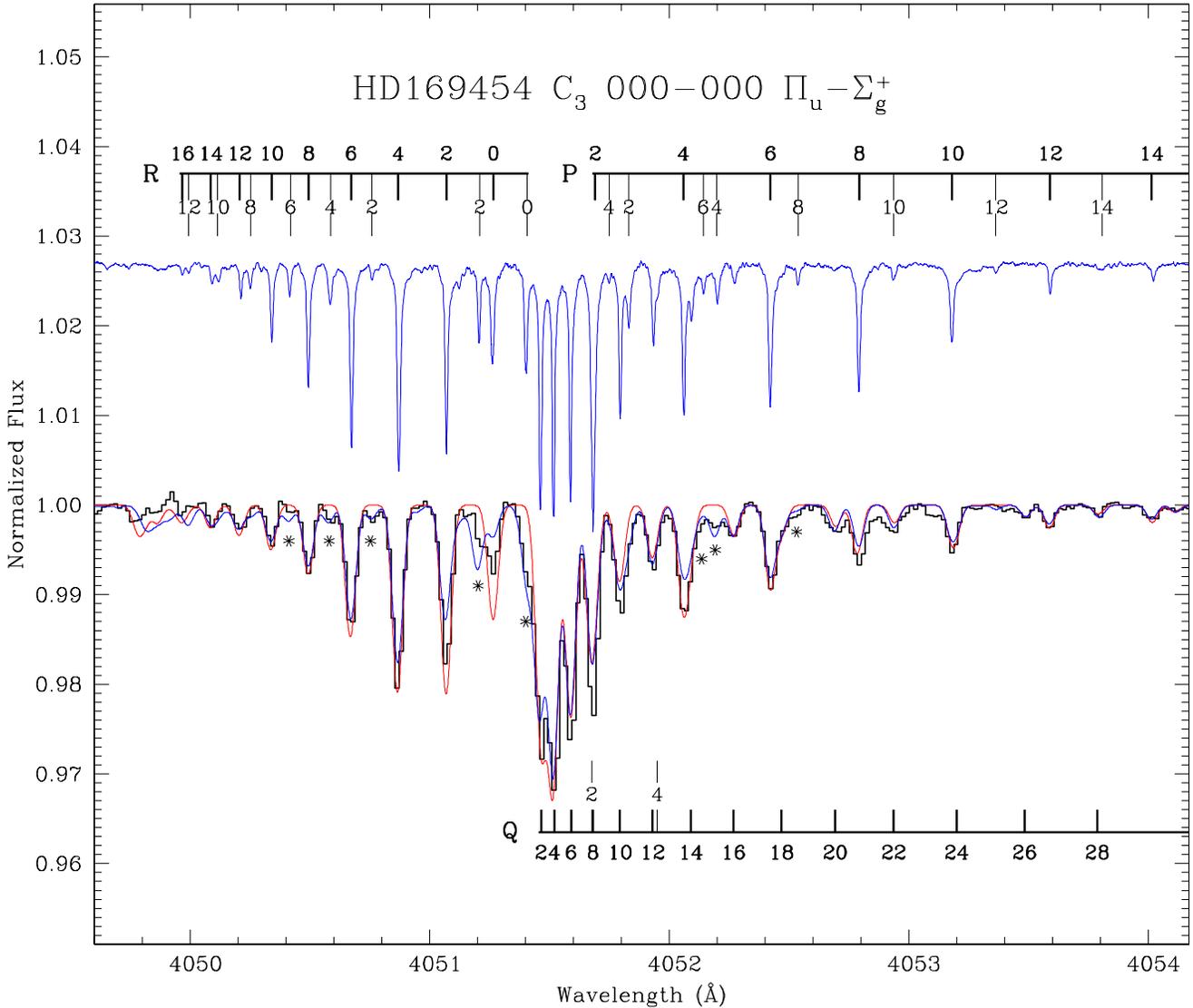}
\caption{
Spectrum of the C$_3$ \~{A}$^{1}\Pi_{u}$ - \~{X}$^{1}\Sigma_{g}^{+}$ 000--000 band.
Upper spectrum: laboratory measurement by cavity ring-down spectroscopy
in a planar plasma jet. Lower spectrum: observation in the sight line to HD~169454.
Positions of rotational lines are tagged with thick lines. Thin lines tag rotational
lines due to perturbing states based on the analysis of~\citet{Zhang2005}.
The astronomical spectrum is overlaid with fitted curves: the red curve
represents a fit using only unperturbed lines; for the blue curve perturber lines
are included in the fit. Note the remaining deviation for the intensity of
the R(0) line in the simulated spectrum (see text).
}
\label{figc3000}
\end{figure*}

The very high S/N of the spectra allowed us to detect seven \~{A}--\~{X}
vibronic C$_3$ bands in addition to the \~{A}--\~{X} origin (000-000)
band in the spectrum of the heavily reddened star: HD~169454.  Four vibronic
bands were also detected in sight lines towards HD~73882 and HD~154368.
These two sight lines are not analysed in detail in this paper.
The resulting spectra for these four bands in all three objects are shown in
Fig.~\ref{fig4bands} and are illustrative for the quality of the
observational data. All targets are heavily reddened and characterized by
strong molecular features.  The individual spectra of identified vibronic
bands in the sight line to HD~169454 are presented in Figs.~\ref{figc3000} -
\ref{figc3120}, alongside with the laboratory spectra, in the order of
excited vibrations: ${\nu_1}{\nu_2^{+/-}}{\nu_3}$ for symmetric stretching,
bending - degenerate-, and asymmetric stretching, respectively, as well as
combination modes.  In this sequence of figures the laboratory spectra are
plotted with blue lines, the astronomical spectra are displayed as black
histograms overlayed with a thin line representing a synthetic spectrum
based on the detailed analysis.  The identification of rotational lines is
indicated by vertical lines.  In the C$_3$ \~{A}$^{1}\Pi_{u}$--
\~{X}$^{1}\Sigma_{g}^{+}$ 000--000 band, see Fig.~\ref{figc3000}, a series
of perturber lines is observed, confirming findings of a laboratory study
by~\citet{Zhang2005}.  These perturber features are marked with
additional thin vertical lines in Fig.~\ref{figc3000}.

For the detailed analysis of the vibronic bands in the sight line to
HD~169454 the spectrum was shifted to the rest wavelength velocity frame
using the CH A-X and B-X lines.  Redshifts as large as 0.85 km s$^{-1}$ were
observed in the case of CN B-X (0,0) lines and 2.8 km s$^{-1}$ in the
case of the Ca~I 4226.7 \AA\ line.  While fitting individual bands some
shifts in wavelengths are necessary, mostly consistent with velocities of
the CN lines.

\subsection{Laboratory experiment}

Laboratory spectra of C$_3$ are recorded in direct absorption using cavity
ring-down laser spectroscopy.  Details of the experimental setup and
experimental procedures can be found elsewhere~\citep{Zhao2011,
Motylewski1999a}.  Briefly, in the current experiment, supersonically
jet-cooled C$_3$ radicals are produced in a pulsed (10 Hz) planar plasma
expansion generated by discharging 0.5\%\ C$_2$H$_2$ diluted in a 1:1
helium/argon gas mixture.  A 3 cm $\times$ 300 $\mu$m slit discharge nozzle
is employed to generate a planar plasma expansion which provides an
essentially Doppler-free environment and a relatively long effective
absorption path length.  The rotational temperature of C$_3$ in the plasma
jet is estimated at 30 K.

Tunable violet light (375 - 410 nm) is generated by frequency-doubling the
near infrared output (750 - 820 nm) of a Nd:YAG-pumped pulsed dye laser
($\sim$6 ns pulse duration).  The bandwidth of the violet laser pulses is
~0.06 cm$^{-1}$.  Since this value exceeds the Doppler width ($<$0.03
cm$^{-1}$) of C$_3$ transitions in the plasma jet, the absolute line
intensities in the recorded spectrum may be underestimated, due to a
specific linewidth effect associated with the cavity ring-down technique.
This effect~\citep{Jongma1995} is small in the case of weak absorptions but
may become pronounced in the case of strong absorptions.  Therefore, in the
present experiment, production densities of C$_3$ in the plasma are
controlled to be not too high, ensuring reliable line intensities for each
individual band, to derive accurate values from the experimental spectra
\citep{Haddad2013}.

Simultaneously with the spectral recordings of C$_3$, transmission fringes
of two etalons (with free spectral ranges of $\sim$20.1 and 7.57 GHz,
respectively) are recorded at the fundamental infrared laser wavelength,
providing relative frequency markers to correct for possible non-linear
wavelength scanning of the dye laser.  The absolute laser frequency is
calibrated by absorption lines of He I or Ar I in the plasma.  In this way,
a wavelength accuracy of better than 0.01~\AA\ is achieved in the final
laboratory spectrum, except for the 02$^+$0 -- 000 band at $\sim$3916 \AA,
around which no strong He I or Ar I lines are found.  The laboratory
spectrum of this 02$^+$0 -- 000 band is calibrated by using the wavelength
of the strongest C$_3$ transition, i.e.  Q(4) line at 3916.05~\AA, in the
astronomical spectrum towards HD~169454.  This yields an absolute wavelength
accuracy of $\sim$0.05 \AA\ for the 02$^+$0 -- 000 band, while the accuracy
of relative line positions within this band is better than 0.01 \AA.

These data sets are used to build a C$_3$ line list, as is discussed
in the next section.

\section{Molecular data}

The vibronic bands in the \~{A}$^{1}\Pi_{u}$-- \~{X}$^{1}\Sigma_{g}^{+}$
electronic system of C$_3$ have been previously investigated in the
laboratory by \citet{Gausset1965, Balfour1994, Tokaryk1997, McCall2003,
Tanabashi2005, Zhang2005, Chen2010}. Most studies were performed at
lower spectral resolution and consequently lower wavelength precision
than the data presented here. The present laboratory spectra, combined with
the previously reported data, allow us to build a highly precise line list
for C$_3$.  Tables~\ref{Table3} -- \ref{Table5} summarize the list of
laboratory line positions, details of which are given in the subsections
below.

\subsection{\~{A}--\~{X} oscillator strength and Franck-Condon factors}

In our analysis we rely on the determination of the oscillator strength from
a measurement of the lifetime of the upper electronic state \~{A}$^1\Pi_u$
corresponding to a value of $f_{el} =0.0246$~\citep{Becker1979}.  With a
calculation of the Franck-Condon factor (FCF) for the 000--000
band~\citep{Radic1977} yielding 0.74, this translates approximately to
$f_{000} = 0.016$, a value previously used in the analyses of interstellar
C$_3$ by \citet{Maier2001}, \citet{Adamkovics2003}, and \citet{Oka2003}.
Computations of FCFs carried out by~\citet{Jungen1980} in an
effective large amplitude formalism lead to 15 percent lower value of
$f_{000} = 0.0146$ used by \citet{Roueff2002}.  Our choice is dictated by
easy comparison to earlier determinations of C$_3$ column densities in sight
line to HD~169454 in light of uncertainty of the FCF value.

Interpretation of the other band intensities requires knowledge
of the Franck-Condon factors for individual vibronic bands.  Calculation of
FCFs is hampered by the occurrence of low-frequency and large-angle bending
modes, as well as by perturbations in the \~{A}$^1\Pi_u$ state, which
affects the \~{A}--\~{X} electronic transition moment as well as the
FCFs.  Computations of FCFs were carried out by~\citet{Jungen1980} in
an effective large amplitude formalism for the transitions involving bending
modes. Calculations performed by~\citet{Radic1977} resulted in
different FCF values.  The intensities in the astronomical spectra
were used to experimentally determine FCFs as well.  For this purpose the
intensities for R(0), R(2), R(4) and R(6) transitions, most easily
identifiable, except in the 120-000 band, were used, and normalized to
the value of $f_{000}=0.016$ of \citet{Radic1977} adopted in this
paper. A summary of available FCFs from the two theoretical studies and the
presently determined experimental values is listed in Table~\ref{Table2}.
The experimentally determined values, that agree in most cases with
the average of both theoretical studies, are used for producing the line
strengths included in the molecular line lists in Tables~\ref{Table3}
-- \ref{Table5}.

\begin{table}
\caption{Franck-Condon factors of the analysed vibronic bands.}
% \label{TableSpec}
\label{Table2}
\begin{tabular}{@{}cllllc}
\hline
Origin  & Band & \multicolumn{3}{c}{FC}& References\\
(\AA) &
& J\&M % \tablefootmark{a}
& R-P  % \tablefootmark{b}
&  this work &   \\
\hline
4051.6 & 000  $\Pi_u$     & 0.594 & 0.741 & 0.741         &  1,2 \\
3880.7 & 100  $\Pi_u$     &       &       & 0.13$\pm$0.02 & 2 \\
3992.8 & 02$^-$0 $\Pi_u$  & 0.087 & 0.170 & 0.14$\pm$0.03 & 3 \\
3916.0 & 02$^+$0 $\Pi_u$  & 0.081 & 0.170 & 0.14$\pm$0.03 & 2 \\
3929.5 & 04$^-$0 $\Pi_u$  & 0.083 & 0.056 & 0.10$\pm$0.03 & 2 \\
3801.7 & 04$^+$0 $\Pi_u$  & 0.048 & 0.056 & 0.10$\pm$0.03 & 4 \\ % Q-branch 0.06, R-banch 0.09
3794.2 & 002 $\Pi_u$      &       &       & 0.08$\pm$0.02 & 4 \\  % average of total 0.07 and R-branch 0.11 determinations
3826.0 & 12$^-$0 $\Pi_u$  &       &       & 0.04$\pm$0.01 & 2,5 \\ % based on eye-estimate
\hline
\end{tabular}

\medskip
{\em J\&M} \citet{Jungen1980}; {\em R-P} \citet{Radic1977};
References: (1) \citet{Zhang2005}; (2) This work;
(3)  \citet{Tokaryk1997}; (4) \citet{Gausset1965}; (5) \citet{Chen2010}.
\end{table}

Because all bands have a $\Pi$ ($K=1$) upper vibronic state and a $\Sigma$ ($K=0$) lower
ground state, the H{\"o}nl-London factors for these
bands, for $\Delta K=1$ (corresponding to both $\Delta l=0$
and $\Delta l=2$) are: $1/2 (J+2)/(2J+1)$, $1/2$, and $1/2 (J-1)/(2J+1)$
for R, Q, and P-branch transitions, respectively. The
heavily perturbed 000-000 band requires a more sophisticated treatment, as is discussed below.

\subsection{\~{A}--\~{X} 000-000 band at 4050 \AA}

The origin band has been analyzed in detail in various laboratory
investigations~\citep{Gausset1965, Zhang2005, McCall2003, Tanabashi2005}.
Earlier indications of a misassigned R(0) transition
were clarified and a number of extra transitions was identified by~\citet{McCall2003}.
\Citet{Zhang2005} proposed an explanation for the existence of extra transitions
via intensity borrowing to two perturbing states lying close to the upper \~{A}$^1\Pi_{u}$ 000 state, and
presented an effective Hamiltonian for the upper \~{A}$^1\Pi_{u}$ 000 state.
The laboratory spectrum for this band, shown in Fig.~\ref{figc3000}, agrees reasonably well with previous
work, and provides more reliable line intensities for the sight line spectrum towards HD~169454,
particularly for the weaker transitions and transitions involving perturbing states.
Using the new laboratory data it was possible to improve the Hamiltonian significantly,
and the perturbed components are now better reproduced.
Note that some perturbed transitions are positively identified in the
astronomical spectra - see transitions marked with an asterisk in Fig.~\ref{figc3000}.
Still, in this perturbation analysis there remains a discrepancy
for the R(0) line,
as is clearly seen in the simulated spectrum, shown in Fig.~\ref{figc3000}.
For the R(0) line there is a nearly equal mixture between singlet and triplet character
with an indication of an additional perturbation for the very lowest $J$-value, not addressed previously.
Oscillator strengths obtained from this new analysis together with
derived column densities are presented in Table~\ref{Table3}.
However, data for the R(0) and P(2) lines of this band
are not used for deriving column densities,
in view of the severe perturbations
in the $J=1$ excited level and the deviations still present in the modeling in this part of the spectrum 
(\emph{cf.} Fig.~\ref{figc3000}).
For $J>20$ the values are taken from~\citet{Tanabashi2005}.

Because of the remaining discrepancies, we propose the following approach to deal with intensities
of perturbed components. Under the assumption of a single electronic transition
moment, perturbed components related to a high vibrational level of the \~{b}$^3\Pi_g$
state borrow intensity from the \~{A}-\~{X} transition.
Summing over equivalent widths of regular and perturbed components,
column densities can be derived for the \~{A}-\~{X} 000-000 band
using line intensities neglecting effects of perturbations.
The corresponding oscillator strengths are then based on the H\"{o}nl-London factors and presented in
Table~\ref{Table4} with wavelengths of regular
components. This approach is finally used in the determination of column densities in Section 4.

\subsection{\~{A}--\~{X} 100-000 band at 3881 \AA}

\begin{figure}
\centering
\includegraphics[width=8.4cm]{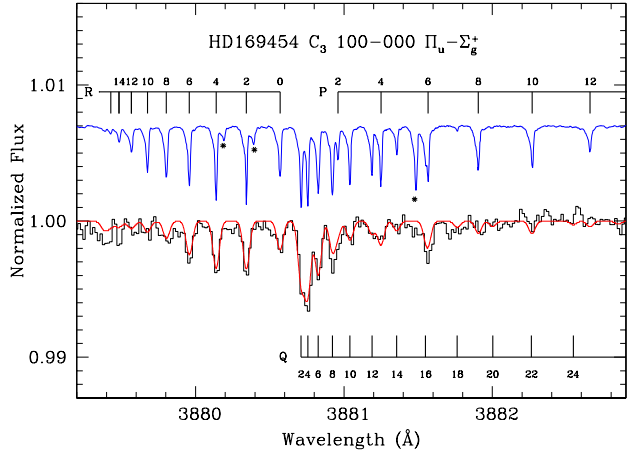}
\caption{
Laboratory spectrum (blue line) and
astronomical spectrum
in the sight line to HD~169454 (black) for the
C$_3$ \~{A}$^{1}\Pi_{u}$ - \~{X}$^{1}\Sigma_{g}^{+}$ 100-000 band.
A synthetic spectrum overplots the observed spectrum with a red line.
Positions of rotational lines are indicated with vertical lines.
Note the presence of additional lines marked by (*) in the laboratory spectrum (see text).
}
\label{figc3100}
\end{figure}

A line list for the 100-000 band is deduced from the present
laboratory spectrum, shown in Fig.~\ref{figc3100}, providing line positions
at an accuracy better than 0.01 \AA.  The line list is extended by
transitions not observed in the laboratory spectrum by the following
procedure.  First, the observed lines are used for the determination of
spectroscopic constants of the upper state, fixing the spectroscopic
constants of the \~{X} ground state to values derived
by~\citet{Tanabashi2005}.  The evaluated spectroscopic constants of the
\~{A} 100 upper state are then used for extending the line list for $J$ up
to 50.  The residuals between experimental and predicted line positions are
in general at the level of 0.01 \AA.  The laboratory spectrum contains three
extra lines, which are not indicative for a perturbation, but rather due to
another unknown species present in the plasma expansion, as cavity ring down
is not mass selective.  These features do not appear in the spectrum of
HD~169454.

\subsection{\~{A}--\~{X} 02$^-$0 -- 000 band at 3993 \AA}

\begin{figure}
\centering
\includegraphics[width=8.4cm]{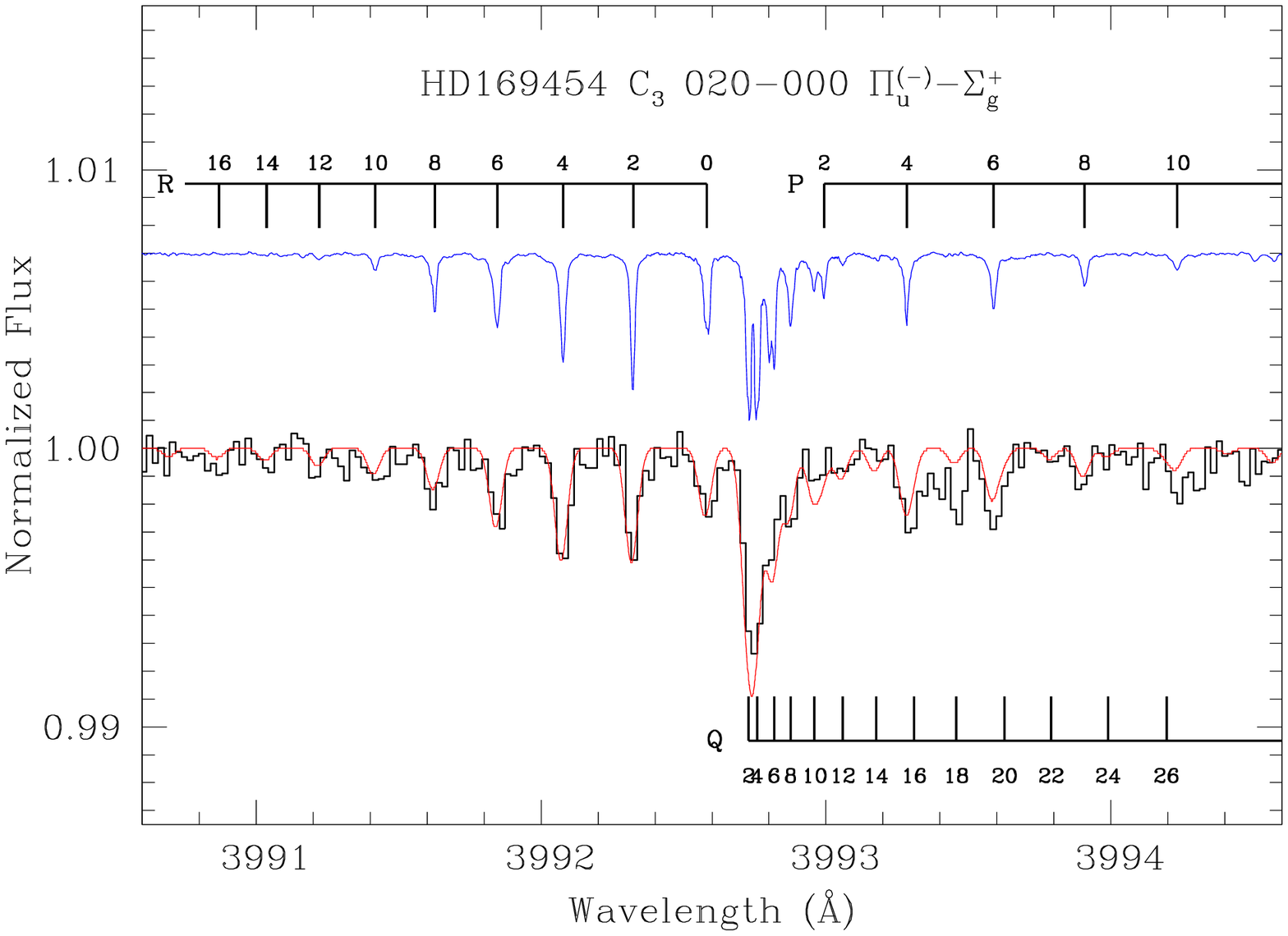}
\caption{
Laboratory spectrum (blue line) and astronomical spectrum
in the sight line to HD~169454 (black) for the
C$_3$ \~{A}$^{1}\Pi_{u}$ - \~{X}$^{1}\Sigma_{g}^{+}$ 02$^-$0 -- 000 band.
}
\label{figc302m0}
\end{figure}

The line positions from \citet{Tokaryk1997},
the most accurate analysis of this band up to now, agree well
with our laboratory data, shown in Fig.~\ref{figc302m0}.
Earlier laboratory work by~\citet{Gausset1965} and \citet{Balfour1994}
suffered from low resolution and most probably from
contamination by 020-020 emission bands, resulting in misassignments of
some rotational lines. Consistency between our spectrum towards HD~169454
and the laboratory spectrum further strengthens the assignments by~\citet{Tokaryk1997}.

\begin{figure}
\centering
\includegraphics[width=8.4cm]{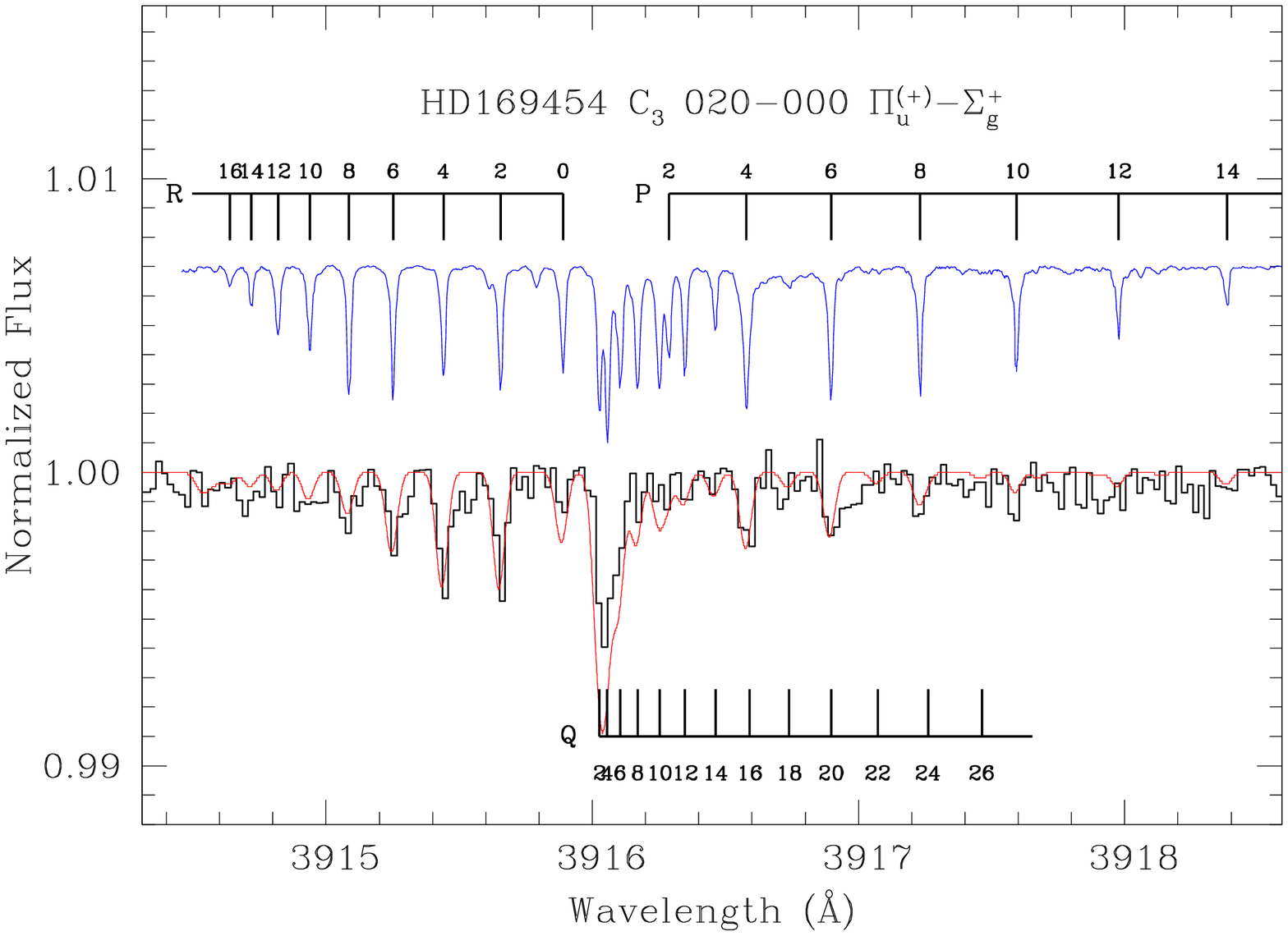}
\caption{
Laboratory spectrum (blue line) and astronomical spectrum in the sight line to HD~169454
(black) for the C$_3$ \~{A}$^{1}\Pi_{u}$ - \~{X}$^{1}\Sigma_{g}^{+}$ 02$^+$0 -- 000 band.
}
\label{figc302p0}
\end{figure}

\begin{figure}
\centering
\includegraphics[width=8.4cm]{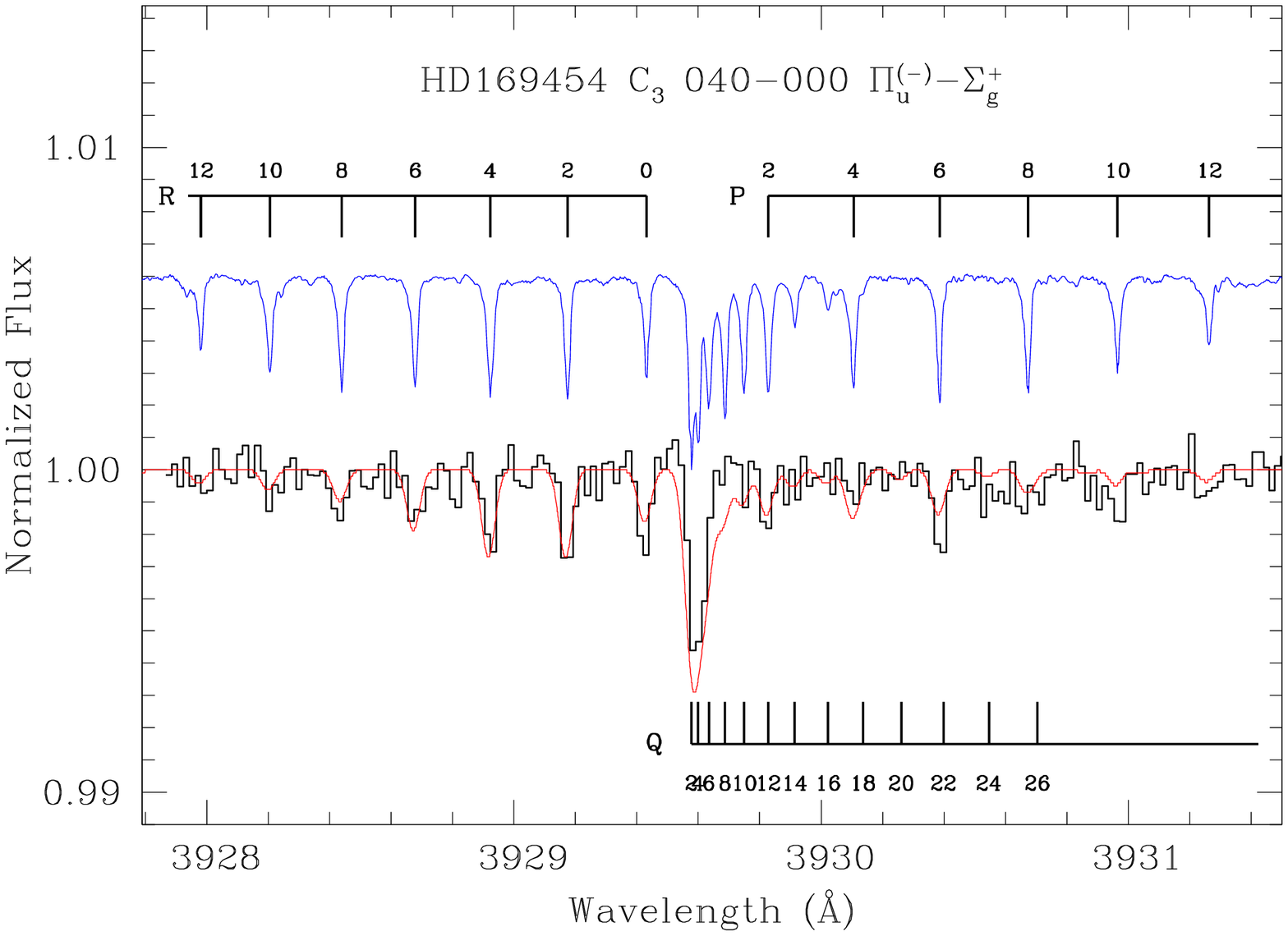}
\caption{
Spectrum in sight line to HD~169454 (black) and laboratory spectrum (blue line) for the
C$_3$ \~{A}$^{1}\Pi_{u}$ - \~{X}$^{1}\Sigma_{g}^{+}$ 04$^-$0 -- 000 band.
}
\label{figc304m0}
\end{figure}

\begin{figure}
\centering
\includegraphics[width=8.4cm]{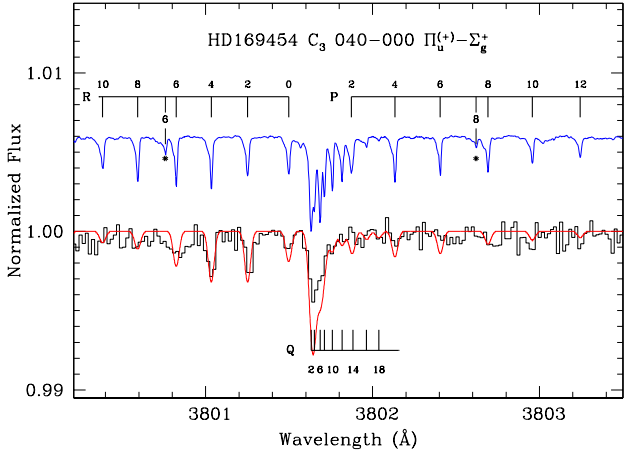}
\caption{
Spectrum in sight line to HD~169454 (black) and laboratory spectrum (blue line) of the
C$_3$ \~{A}$^{1}\Pi_{u}$ - \~{X}$^{1}\Sigma_{g}^{+}$ 04$^+$0 -- 000 band.
Perturber lines are marked with (*); see text.
}
\label{figc304p0}
\end{figure}

\begin{figure}
\centering
\includegraphics[width=8.4cm]{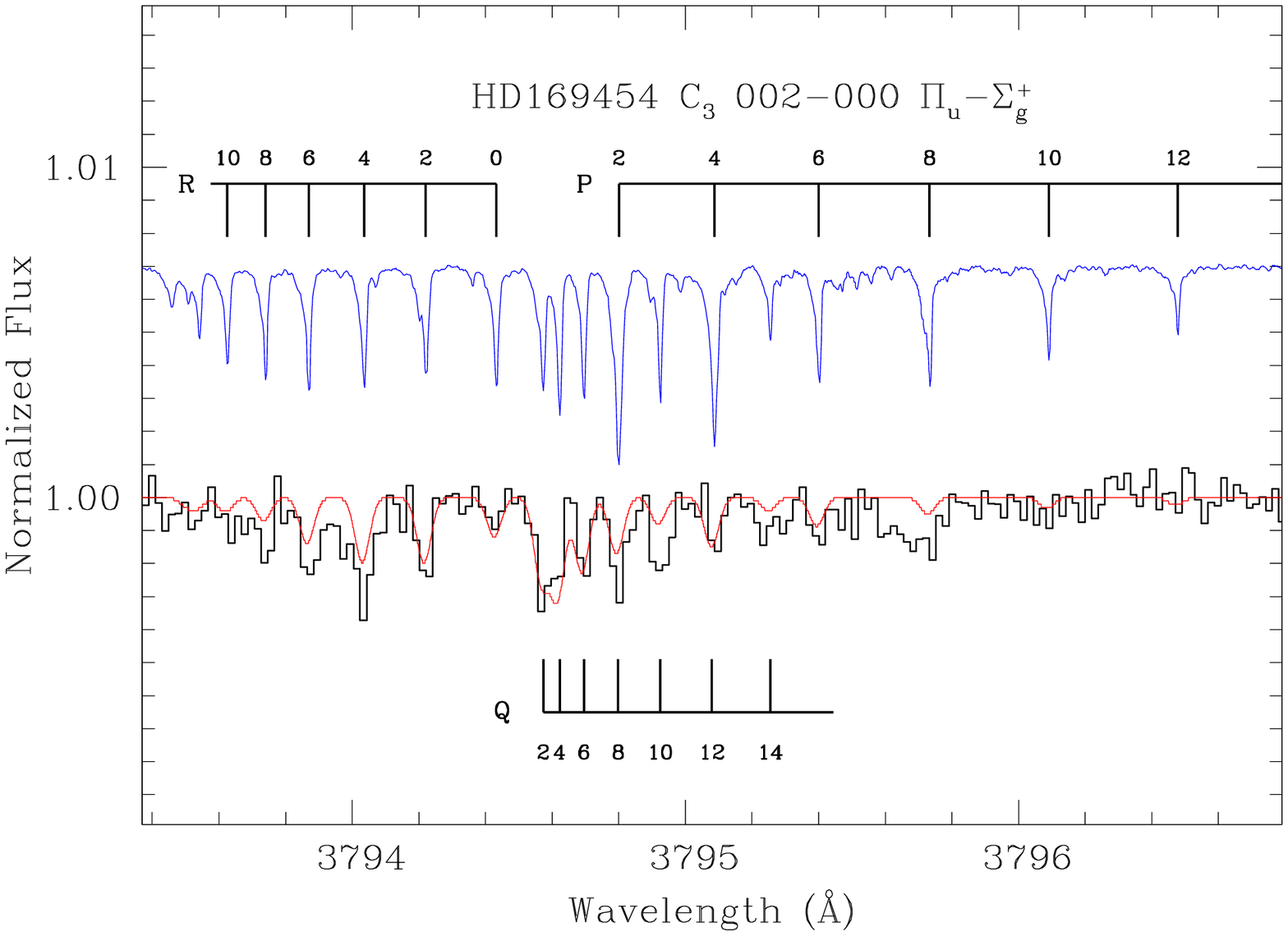}
\caption{
Spectrum in sight line to HD~169454 (black) and laboratory spectrum (blue line) of the
C$_3$ \~{A}$^{1}\Pi_{u}$ - \~{X}$^{1}\Sigma_{g}^{+}$ 002-000 band.
}
\label{figc3002}
\end{figure}

\begin{figure}
\centering
\includegraphics[width=8.4cm]{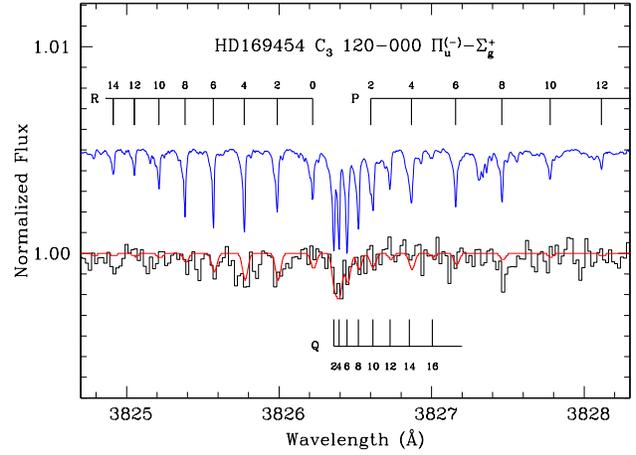}
\caption{
Spectrum in sight line to HD~169454 (black) and laboratory spectrum (blue line) of the
C$_3$ \~{A}$^{1}\Pi_{u}$ - \~{X}$^{1}\Sigma_{g}^{+}$ 120 -- 000 band.
}
\label{figc3120}
\end{figure}

\subsection{\~{A}--\~{X} 02$^+$0 --- 000 band at 3916 \AA}

Entries for the line list are taken from the present laboratory spectrum,
shown in Fig.~\ref{figc302p0}, and extended to higher-$J$ transitions based
on the deduced molecular constants.  As the absolute wavelength
calibration relies on the observational spectrum towards HD~169454, the
wavelength positions in this band are of a lower accuracy, at the level of
$\sim 0.05$ \AA.

\subsection{\~{A}--\~{X} 04$^-$0 -- 000 band at 3930\AA}

The line list is extracted from the present laboratory data,
shown in Fig.~\ref{figc304m0}, and extended to higher-$J$ transitions.
Previously reported data by \citet{Gausset1965} and \citet{Balfour1994}
do not reproduce the observed spectrum in a consistent manner,
likely for misassign as explained in Section 3.4.

\subsection{\~{A}--\~{X} 04$^+$0 -- 000 band at 3802 \AA}

The line list is extracted from the present laboratory data, shown in
Fig.~\ref{figc304p0}.  Extra lines in the laboratory spectrum marked with an
asterisk (*) are associated with perturbation features; based on combination
differences these extra lines are assigned as R(6) and P(8) lines.

\subsection{\~{A}--\~{X} 002-000 band at 3794 \AA}

The line list is extracted from the laboratory data shown in
Fig.~\ref{figc3002}.  Spectroscopic constants for the upper state are
derived, yielding more accurate values than in \citet{Chen2010}.  Large
residuals of $\sim$0.05 \AA\ are found indicating perturbations in the upper
state for $J>14$ transitions.  Therefore, the line list of this band
is not extended beyond this $J$ value.

\subsection{\~{A}--\~{X} 120 -- 000 band at 3826 \AA}

The line list for this band is extracted from the laboratory data.  Due to
its weakness, this band is barely visible towards HD~169454. However
the detection of this band, based on the low-$J$ Q branch lines clearly
visible in Fig.~\ref{figc3120}, is certain.  The quality of the data
is not of sufficient accuracy to determine line strengths or column
densities.

\vspace{0.5cm}

The line lists providing the C$_3$ molecular data are presented in
Table~\ref{Table3} for the \~{A}-\~{X} 000-000 band
taking into account effect of perturbations,
in Table~\ref{Table4} for regular transitions only,
and in Table~\ref{Table5} for the vibronic bands probing excited
vibrations in the upper state, with exception of the 120-000 band system.
For a full list of laboratory line positions of the \~{A}-\~{X} 000-000 band
and a more detailed account of the perturbation analysis we refer to \cite{Haddad2013}.

\begin{table}
\caption{The C$_3$ line list for the \~{A}-\~{X} 000-000 band as obtained from the present analysis.
Perturber lines are marked with an asterisk (*).}
% \label{TableLinesC3000}
\label{Table3}
\centering
\begin{tabular}{llrlll}
\hline\hline
Wavelength & Line  &  $f_{JJ}$   &   EW    & N$_{col}$ \\
   (\AA)   &       & ($\times$10$^3$) &  (m\AA) & (10$^{12}$ cm$^{-2}$) \\
\hline
4050.075 & R(14)  &    3.49  & 0.10$\pm$0.03 & 0.20$\pm$0.06\\
4050.191 & R(12)  &    3.58  & 0.16$\pm$0.03 & 0.31$\pm$0.06\\
4050.327 & R(10)  &    3.73  & 0.21$\pm$0.03 & 0.39$\pm$0.06\\
4050.401*& R(6)   &    0.58  & 0.03$\pm$0.02 & 0.36$\pm$0.24\\
4050.484 & R(8)   &    3.95  & 0.37$\pm$0.04 & 0.65$\pm$0.07\\
4050.567*& R(4)   &    0.49  & 0.08$\pm$0.03 & 1.12$\pm$0.42\\
4050.661 & R(6)   &    4.21  & 0.74$\pm$0.07 & 1.21$\pm$0.11\\
4050.746*& R(2)   &    0.45  & 0.08$\pm$0.04 & 1.22$\pm$0.61\\
4050.857 & R(4)   &    4.44  & 1.06$\pm$0.11 & 1.64$\pm$0.17\\
4051.055 & R(2)   &    3.78  & 0.96$\pm$0.10 & 1.75$\pm$0.18\\
4051.190*& R(2)   &    2.08  & 0.22$\pm$0.05 & 0.73$\pm$0.17\\
% 4051.255 & R(0)   &    4.22  & 0.41$\pm$0.08 & 0.67$\pm$0.13\\
% 4051.396*& R(0)   &   10.58  & 0.36$\pm$0.12 & 0.23$\pm$0.08\\
4051.255 & R(0)   &    4.22  & 0.41$\pm$0.08 & \\
4051.396*& R(0)   &   10.58  & 0.36$\pm$0.12 & \\
4051.448 & Q(2)   &    6.87  & 1.35$\pm$0.26 & 1.35$\pm$0.26\\
4051.506 & Q(4)   &    7.70  & 1.76$\pm$0.36 & 1.57$\pm$0.32\\
4051.578 & Q(6)   &    7.89  & 1.44$\pm$0.28 & 1.26$\pm$0.24\\
4051.782 & Q(10)  &    7.96  & 0.64$\pm$0.14 & 0.55$\pm$0.12\\
% 4051.820 & P(2)   &    1.06  & 0.14$\pm$0.07 & 0.91$\pm$0.45\\
4051.820 & P(2)   &    1.06  & 0.14$\pm$0.07 & \\
4051.918 & Q(12)  &    7.97  & 0.38$\pm$0.06 & 0.33$\pm$0.05\\
4052.045 & P(4)   &    1.57  & 0.56$\pm$0.14 & 2.45$\pm$0.61\\
4052.077 & Q(14)  &    7.98  & 0.27$\pm$0.07 & 0.23$\pm$0.06\\
4052.122*& P(6)   &    0.28  & 0.09$\pm$0.04 & 2.21$\pm$0.98\\
4052.180*& P(4)   &    0.86  & 0.15$\pm$0.04 & 1.20$\pm$0.32\\
4052.257 & Q(16)  &    7.98  & 0.20$\pm$0.03 & 0.17$\pm$0.03\\
4052.412 & P(6)   &    2.56  & 0.49$\pm$0.08 & 1.32$\pm$0.22\\
4052.459 & Q(18)  &    7.99  & 0.13$\pm$0.08 & 0.11$\pm$0.07\\
4052.521*& P(8)   &    0.39  & 0.06$\pm$0.03 & 1.06$\pm$0.53\\
4052.683 & Q(20)  &    7.99  & 0.20$\pm$0.08 & 0.17$\pm$0.07\\
4052.782 & P(8)   &    2.82  & 0.40$\pm$0.08 & 0.98$\pm$0.20\\
4053.169 & P(10)  &    2.87  & 0.30$\pm$0.04 & 0.72$\pm$0.10\\
4053.479 & Q(26)  &    7.99  & 0.08$\pm$0.03 & 0.07$\pm$0.03\\
4053.577 & P(12)  &    2.87  & 0.14$\pm$0.05 & 0.34$\pm$0.12\\
4053.783 & Q(28)  &    8.00  & 0.05$\pm$0.03 & 0.04$\pm$0.03\\
4054.005 & P(14)  &    2.86  & 0.09$\pm$0.03 & 0.22$\pm$0.07\\
\hline\hline
\end{tabular}
\end{table}

\begin{table}
\caption{The C$_3$ line list for the \~{A}-\~{X} 000-000 band as obtained from
unperturbed intensities only.}
% \label{TableLinesC3000nopert}
\label{Table4}
\centering
\begin{tabular}{llrlll}
\hline\hline
Wavelength$^a$ & Line  &  $f_{JJ}$   &   EW    & N$_{col}$ \\
   (\AA)   &       & ($\times$10$^3$) &  (m\AA) & (10$^{12}$ cm$^{-2}$) \\
\hline
%
% 000 -- 000  & & & & & \\
%
4050.075 & R(14) &    4.41  &  0.10$\pm$0.03 & 0.16$\pm$0.05\\
4050.191 & R(12) &    4.48  &  0.16$\pm$0.03 & 0.25$\pm$0.05\\
4050.327 & R(10) &    4.57  &  0.21$\pm$0.03 & 0.32$\pm$0.05\\
4050.484 & R(8)  &    4.72  &  0.37$\pm$0.04 & 0.54$\pm$0.06\\
4050.661 & R(6)  &    4.92  &  0.77$\pm$0.09 & 1.08$\pm$0.13\\
4050.857 & R(4)  &    5.33  &  1.14$\pm$0.14 & 1.47$\pm$0.18\\
4051.055 & R(2)  &    6.40  &  1.26$\pm$0.19 & 1.36$\pm$0.20\\
4051.396 & R(0)  &   16.00  &  0.77$\pm$0.20 & 0.33$\pm$0.09\\
4051.448 & Q(2)  &    8.00  &  1.35$\pm$0.26 & 1.16$\pm$0.22\\
4051.506 & Q(4)  &    8.00  &  1.76$\pm$0.36 & 1.51$\pm$0.31\\
4051.578 & Q(6)  &    8.00  &  1.44$\pm$0.28 & 1.24$\pm$0.24\\
4051.782 & Q(10) &    8.00  &  0.64$\pm$0.14 & 0.55$\pm$0.12\\
4051.918 & Q(12) &    8.00  &  0.38$\pm$0.06 & 0.33$\pm$0.05\\
4052.045 & P(4)  &    2.67  &  0.71$\pm$0.18 & 1.83$\pm$0.46\\
4052.077 & Q(14) &    8.00  &  0.27$\pm$0.07 & 0.23$\pm$0.06\\
4052.257 & Q(16) &    8.00  &  0.20$\pm$0.03 & 0.17$\pm$0.03\\
4052.412 & P(6)  &    2.56  &  0.58$\pm$0.12 & 1.56$\pm$0.32\\
4052.459 & Q(18) &    8.00  &  0.13$\pm$0.08 & 0.11$\pm$0.07\\
4052.683 & Q(20) &    8.00  &  0.20$\pm$0.08 & 0.17$\pm$0.07\\
4052.782 & P(8)  &    3.29  &  0.46$\pm$0.11 & 0.96$\pm$0.23\\
4053.169 & P(10) &    3.43  &  0.30$\pm$0.04 & 0.60$\pm$0.08\\
4053.479 & Q(26) &    8.00  &  0.08$\pm$0.03 & 0.07$\pm$0.03\\
4053.577 & P(12) &    3.52  &  0.14$\pm$0.05 & 0.27$\pm$0.10\\
4053.783 & Q(28) &    8.00  &  0.05$\pm$0.03 & 0.04$\pm$0.03\\
4054.005 & P(14) &    3.59  &  0.09$\pm$0.03 & 0.17$\pm$0.06\\
\hline\hline
\end{tabular}
\\
$^a$ wavelengths of the regular components
\end{table}

\begin{table}
\caption{The C$_3$ line list for transitions to excited vibrational levels in the \~{A} state.}
% \label{TableLinesC3000cd}
\label{Table5}
\centering
\begin{tabular}{llrlll}
\hline\hline
Wavelength & Line  &  $f_{JJ}$   &   EW    & N$_{col}$ \\
   (\AA)   &       & ($\times$10$^3$) &  (m\AA) & (10$^{12}$ cm$^{-2}$) \\
\hline
100-000  & & & & & \\
3879.797   & R(8)   &    0.83  & 0.09$\pm$0.04 & 0.81$\pm$0.36\\
3879.952   & R(6)   &    0.87  & 0.16$\pm$0.03 & 1.38$\pm$0.26\\
3880.132   & R(4)   &    0.93  & 0.18$\pm$0.03 & 1.45$\pm$0.24\\
3880.336   & R(2)   &    1.13  & 0.18$\pm$0.03 & 1.20$\pm$0.20\\
3880.563   & R(0)   &    2.82  & 0.11$\pm$0.03 & 0.29$\pm$0.08\\
3880.705   & Q(2)   &    1.41  & 0.21$\pm$0.03 & 1.12$\pm$0.16\\
3880.750   & Q(4)   &    1.41  & 0.33$\pm$0.03 & 1.76$\pm$0.16\\
3880.820   & Q(6)   &    1.41  & 0.20$\pm$0.03 & 1.06$\pm$0.16\\
3880.916   & Q(8)   &    1.41  & 0.17$\pm$0.03 & 0.90$\pm$0.16\\
3880.953   & P(2)   &    0.28  & 0.08$\pm$0.03 & 2.14$\pm$0.80\\
3881.034   & Q(10)  &    1.41  & 0.07$\pm$0.03 & 0.37$\pm$0.16\\
3881.182   & Q(12)  &    1.41  & 0.06$\pm$0.03 & 0.32$\pm$0.16\\
3881.243   & P(4)   &    0.47  & 0.07$\pm$0.03 & 1.12$\pm$0.48\\
02$^-$0 -- 000   & & & & \\
3991.620   & R(8)   &    0.93  & 0.12$\pm$0.03 & 0.92$\pm$0.23\\
3991.840   & R(6)   &    0.97  & 0.16$\pm$0.03 & 1.17$\pm$0.22\\
3992.070   & R(4)   &    1.05  & 0.24$\pm$0.03 & 1.62$\pm$0.20\\
3992.317   & R(2)   &    1.26  & 0.20$\pm$0.04 & 1.13$\pm$0.23\\
3992.575   & R(0)   &    3.15  & 0.14$\pm$0.03 & 0.32$\pm$0.07\\
02$^+$0-000  &  & & & \\
3915.080   & R(8)   &    0.92  &  0.11$\pm$0.03 & 0.88$\pm$0.24\\
3915.246   & R(6)   &    0.96  &  0.15$\pm$0.03 & 1.15$\pm$0.23\\
3915.435   & R(4)   &    1.04  &  0.22$\pm$0.03 & 1.56$\pm$0.21\\
3915.649   & R(2)   &    1.25  &  0.21$\pm$0.03 & 1.24$\pm$0.18\\
3915.884   & R(0)   &    3.13  &  0.07$\pm$0.03 & 0.16$\pm$0.07\\
3916.573   & P(4)   &    0.53  &  0.14$\pm$0.03 & 1.95$\pm$0.42\\
3916.890   & P(6)   &    0.61  &  0.13$\pm$0.04 & 1.57$\pm$0.48\\
3917.225   & P(8)   &    0.64  &  0.09$\pm$0.03 & 1.04$\pm$0.35\\
04$^-$0 -- 000  &    & \\
3928.433   & R(8)   &    0.62  &  0.08$\pm$0.03 & 0.94$\pm$0.35\\
3928.672   & R(6)   &    0.65  &  0.10$\pm$0.04 & 1.13$\pm$0.45\\
3928.917   & R(4)   &    0.71  &  0.12$\pm$0.03 & 1.24$\pm$0.31\\
3929.168   & R(2)   &    0.85  &  0.15$\pm$0.04 & 1.29$\pm$0.34\\
3929.425   & R(0)   &    2.11  &  0.12$\pm$0.03 & 0.42$\pm$0.10\\
04$^+$0 -- 000  &  &  & \\
3801.021   & R(4)  &   0.72   & 0.15$\pm$0.02 & 1.63$\pm$0.22\\
3801.239   & R(2)  &   0.87   & 0.13$\pm$0.03 & 1.17$\pm$0.27\\
3801.485   & R(0)  &   2.17   & 0.06$\pm$0.03 & 0.22$\pm$0.11\\
002 -- 000  & & & & & \\
3793.734   & R(8)  &    0.48  &  0.10$\pm$0.04 & 1.64$\pm$0.65\\
3793.864   & R(6)  &    0.51  &  0.15$\pm$0.04 & 2.31$\pm$0.62\\
3794.030   & R(4)  &    0.55  &  0.20$\pm$0.10 & 2.85$\pm$1.43\\
3794.215   & R(2)  &    0.66  &  0.13$\pm$0.03 & 1.55$\pm$0.36\\
3794.426   & R(0)  &    1.66  &  0.04$\pm$0.03 & 0.19$\pm$0.14\\
3794.567   & Q(2)  &    0.84  &  0.16$\pm$0.04 & 1.49$\pm$0.37\\
3794.616   & Q(4)  &    0.84  &  0.10$\pm$0.03 & 0.93$\pm$0.28\\
3794.689   & Q(6)  &    0.84  &  0.11$\pm$0.03 & 1.03$\pm$0.28\\
3794.918   & Q(10) &    0.84  &  0.14$\pm$0.07 & 1.31$\pm$0.65\\
3795.081   & P(4)  &    0.27  &  0.07$\pm$0.03 & 2.03$\pm$0.87\\
\hline\hline
\end{tabular}
\end{table}

\begin{figure}
\centering
\includegraphics[width=83mm]{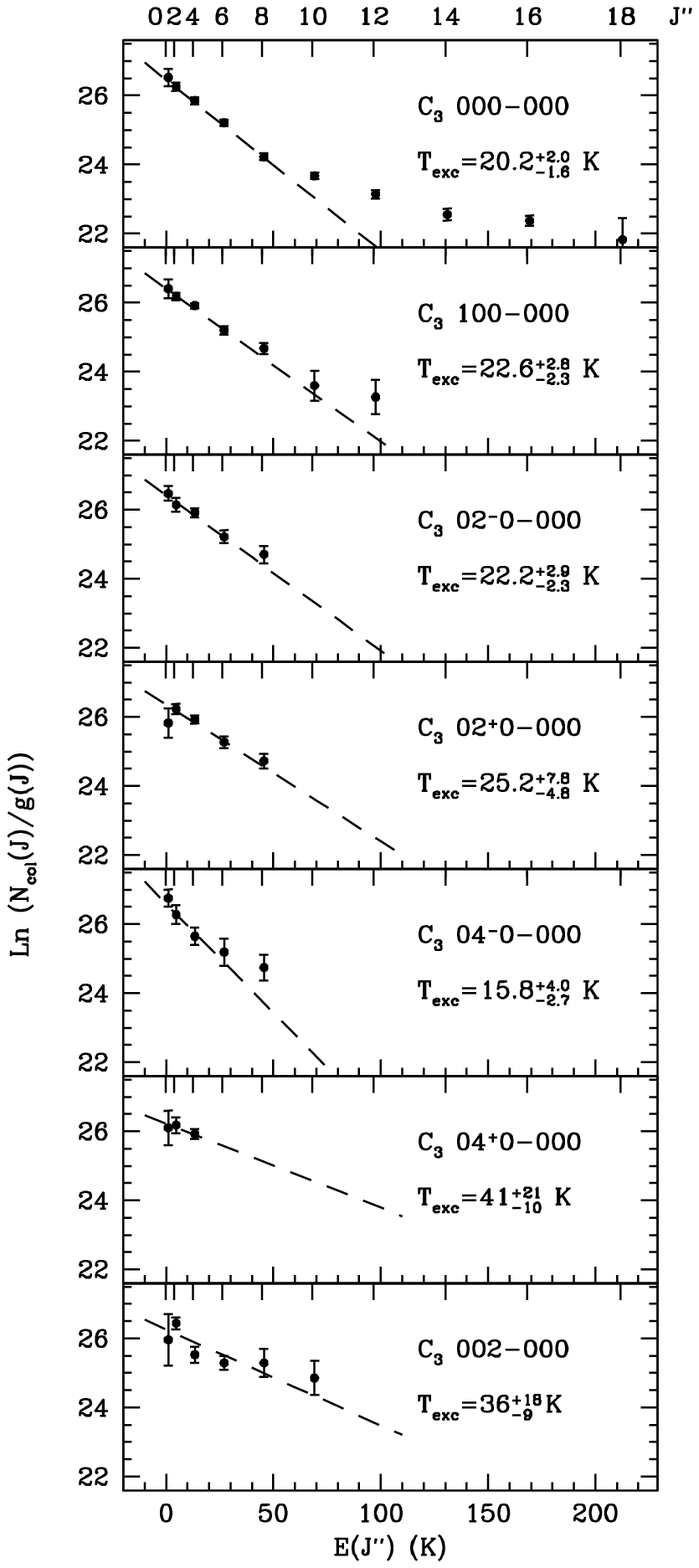}
\caption{Information on column densities in HD~169454, as derived from seven vibronic bands in the
\~{A}-\~{X} system of C$_3$, presented as Boltzmann plots. Note that the limited energy
range of the diagram chosen for better illustration of all seven bands resulted in absence
of the higher rotational levels in the 000-000 band.
}
\label{fig10}
\end{figure}

\section{Data analysis; column densities}

Equivalent widths of rotationally resolved lines in vibronic bands
of the C$_3$ \~{A}-\~{X} system have been measured using DECH and independently with
the \emph{deblend} task of the IRAF package.
Errors of equivalent widths were based on a noise model with
constant Gaussian width taken from a measured
S/N estimated at 2000.
When tracing of the continuum was dubious, a systematic error was added.
Further, in case of blends, or broad absorptions underlying spectral lines,
the full width at half maximum (FWHM) of a Gaussian profile was fixed to the value of the instrumental width
(0.051 \AA) for the astronomical observations (corresponding to $R=80,000$).
This value for the FWHM of the instrumental profile is confirmed by
the best fits to clean absorbing lines.
The C$_{3}$ line positions  were found to be systematically redshifted
by a value comparable or smaller than the accuracy of the laboratory measurements, 0.01
\AA, corresponding to a maximal redshift of 0.8 km\,s$^{-1}$.

Observations of C$_2$ in the sight line to HD~169454 with the Ultra High
Resolution Facility~\citep{Crawford1997} indicated the existence of at
least two velocity components present in the intervening diffuse cloud.
Yet, the components are barely resolved and a possible splitting has been
ignored in our analysis.  Since the C$_3$ lines in the present study are
relatively weak, W$_\lambda \leq 2$ m\AA, column densities of C$_3$ can be
deduced assuming the cloud to be optically thin, so that the total column
density is equal to the sum of those of the velocity components.  This
assumption may influence the final excitation temperatures but does not
affect the observed column densities.

\begin{figure}
\centering
\includegraphics[width=85mm]{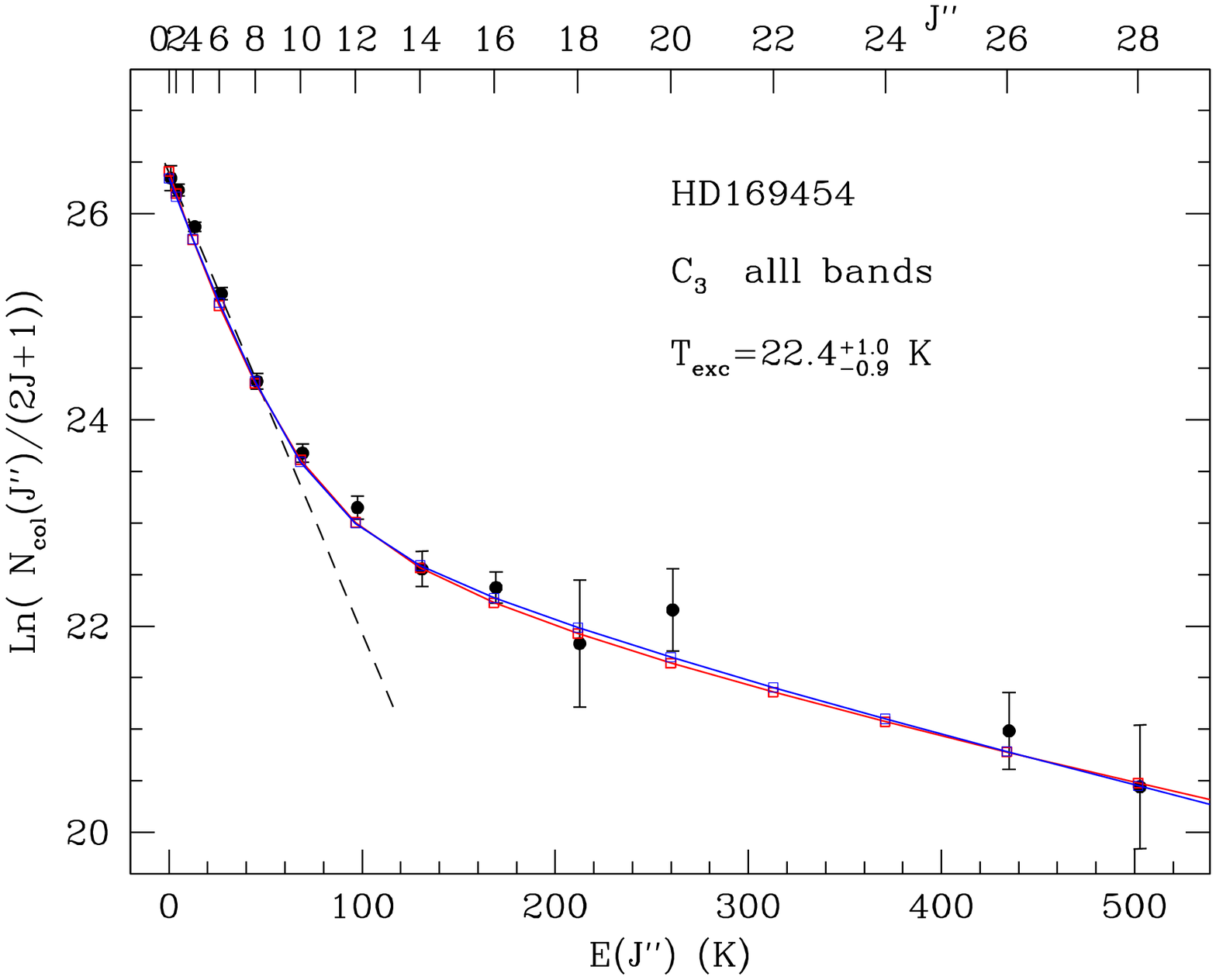}
\caption{Column density distribution of C$_3$ {X}$^{1}\Sigma_{g}^{+}$ 000 rotational
levels towards HD~169454 derived from observations of the \~{A}-\~{X}
bands.
The dashed line shows the thermal population at 22.4 K obtained from the weighted
linear fit to levels $J=0,2,4,6$.
The solid red and blue lines
% {\bf and dashed green line}
show the predictions of the excitation models described in the text.
}
\label{figrotdiagram}
\end{figure}

Column densities of the \~{X} ground state levels for each quantum number
$J$ are derived from the equivalent widths and based on the line lists of
Tables~\ref{Table4} and \ref{Table5}. Note that the equivalent widths
in Table~\ref{Table4} are sum of regular and perturbed components available
from Table~\ref{Table3}. The resulting column densities are listed in
Table~\ref{TableNcol} for each band separately.  The last column contains
the weighted average of all analysed bands.  The one-but-last row of
Table~\ref{TableNcol} shows column densities summed over the observed
transitions in each separate band. The derived column densities, listed in
Table~\ref{TableNcol}, are presented in the form of Boltzmann plots for
individual bands in Fig.~\ref{fig10}, up to $J=18$.  An extended Boltzmann
plot for the weighted average column densities of the analysed
\~{A}-\~{X} bands, including higher $J$ levels solely dependent on the
\~{A}-\~{X} 000-000 band, is presented in Fig.~\ref{figrotdiagram}.  In such
plots the negative inverse of the slope of the straight line is equal to the
excitation temperature.  For each band we show the least square fit for the
lowest $J=0,2,4,6$ rotational levels and the resulting value of the
excitation temperature.  The distribution of the column densities shows a
characteristic "two-temperature" behaviour, as was also found in previous
studies (\emph{e.g.}, \citet{Maier2001}, see their Fig. 5).

\begin{table*}
\caption{Column densities for C$_3$ quantum states in individual vibronic bands
and weighted average of all vibronic bands.}
\label{TableNcol}
\centering
\begin{tabular}{rrrrrrrrrr}
\hline\hline
$J$ & Energy      & \multicolumn{8}{c}{N$_{col}$} \\
    & (cm$^{-1}$) & \multicolumn{8}{c}{(10$^{12}$ cm$^{-2}$)} \\
    &           &  000 -- 000    &  100 -- 000     &  02$^-$0 -- 000 &  02$^+$0 -- 000 &  04$^-$0 -- 000 &  04$^+$0 -- 000 &  002 -- 000  &
all bands$^a$ \\
\hline
   0 &    0.0000 &  0.33$\pm$0.09 &  0.29$\pm$0.08  &  0.32$\pm$0.07  &  0.16$\pm$0.07  &  0.42$\pm$0.10  &  0.22$\pm$0.11  &  0.19$\pm$0.14 & 0.28$\pm$0.03 \\
   2 &    2.5835 &  1.27$\pm$0.15 &  1.17$\pm$0.12  &  1.13$\pm$0.23  &  1.24$\pm$0.18  &  1.29$\pm$0.34  &  1.17$\pm$0.27  &  1.52$\pm$0.26 & 1.23$\pm$0.07 \\
   4 &    8.6112 &  1.52$\pm$0.15 &  1.62$\pm$0.13  &  1.62$\pm$0.20  &  1.64$\pm$0.19  &  1.24$\pm$0.31  &  1.63$\pm$0.22  &  1.10$\pm$0.26 & 1.55$\pm$0.07 \\
   6 &   18.0822 &  1.16$\pm$0.11 &  1.15$\pm$0.14  &  1.17$\pm$0.22  &  1.23$\pm$0.21  &  1.13$\pm$0.45  &                 &  1.25$\pm$0.26 & 1.17$\pm$0.07 \\
   8 &   30.9950 &  0.57$\pm$0.06 &  0.89$\pm$0.15  &  0.92$\pm$0.23  &  0.93$\pm$0.20  &  0.94$\pm$0.35  &                 &  1.64$\pm$0.65 & 0.65$\pm$0.05 \\
  10 &   47.3475 &  0.40$\pm$0.04 &  0.37$\pm$0.16  &                 &                 &                 &                 &  1.31$\pm$0.65 & 0.40$\pm$0.04 \\
  12 &   67.1372 &  0.28$\pm$0.03 &  0.32$\pm$0.16  &                 &                 &                 &                 &                & 0.28$\pm$0.03 \\
  14 &   90.3612 &  0.18$\pm$0.03 &                 &                 &                 &                 &                 &                & 0.18$\pm$0.03 \\
  16 &  117.0160 &  0.17$\pm$0.03 &                 &                 &                 &                 &                 &                & 0.17$\pm$0.03 \\
  18 &  147.0976 &  0.11$\pm$0.07 &                 &                 &                 &                 &                 &                & 0.11$\pm$0.07 \\
  20 &  180.6020 &  0.17$\pm$0.07 &                 &                 &                 &                 &                 &                & 0.17$\pm$0.07 \\
  22 &  217.5244 &                &                 &                 &                 &                 &                 &                &  \\
  24 &  257.8600 &                &                 &                 &                 &                 &                 &                &  \\
  26 &  301.6039 &  0.07$\pm$0.03 &                 &                 &                 &                 &                 &                & 0.07$\pm$0.03 \\
  28 &  348.7507 &  0.04$\pm$0.03 &                 &                 &                 &                 &                 &                & 0.04$\pm$0.03 \\
     &           &                &                 &                 &                 &                 &                 &                &  \\
\multicolumn{2}{l}{Obs. N$_{col}$}
                 &  \emph{6.27$\pm$0.28} &  \emph{5.81$\pm$0.36}  &
\emph{5.15$\pm$0.44}  &  \emph{5.20$\pm$0.39}  &  \emph{5.02$\pm$0.74}  &
\emph{3.01$\pm$0.36}  &  \emph{7.00$\pm$1.04} & \emph{6.31$\pm$0.18} \\
\multicolumn{2}{l}{Tot. N$_{col}$$^a$}
                 &  6.57$\pm$0.29 &                 &                 &                 &                 &                 &                & 6.61$\pm$0.19 \\
\hline
\end{tabular}
\\
$^a$ see text for explanation
\end{table*}

To assess the quality of the analysis and line lists we have performed a
simulation of all observed bands based on the obtained parameters for the
column densities and the excitation temperature.  Synthetic spectra
are computed, assuming an instrumental FWHM of 0.051\,\AA, and
included in Figs.~\ref{figc3000}-\ref{figc3120} as a thin red line.
Inspection of the eight vibronic bands observed toward HD~169454 (see
Figs.~\ref{figc3000}-\ref{figc3120}) shows that intensities in the Q-branch
are slightly overestimated in spectra of the bending modes.  When using line
lists compiled from literature data, this caused a serious problem,
particularly in the case of the 04$^+$0-000 band, where the best fit to the
R branch overestimated the Q band intensity at least twice.  With the
newly composed line lists (Tables~\ref{Table3} and
\ref{Table5}) these discrepancies vanished, suggesting that the
origin of the inconsistencies were in erroneous assignments of the lines,
and/or improper treatment of perturbations in the spectra.

The excitation temperature of the lowest levels is expected to be close to
the gas kinetic temperature in analogy to the excitation model for C$_2$
\citep{Dishoeck1982}.  The information of the weighted average column
densities of all bands is used to derive an accurate excitation
temperature.  The low-$J$ rotational lines of the \~A-\~X 000-000 band
are strongly perturbed, but using the procedure of summation of equivalent
widths of regular and perturbed transitions we believe to have reduced
significantly the effect of unknown intensities of individual components.
The excitation temperature of the lowest $J=0, 2$, $4$, and $6$ levels
determined from the best fit is found to be $22.4^{+1.0}_{-0.9}$ K.
The excitation temperature of the high-$J$ tail is $187^{+25}_{-19}$ K
(see also Fig.~\ref{figrotdiagram}).

For the determination of the total column density an estimate for the
population of the unobserved levels must be invoked.  This is
accomplished by fitting the observed average column densities with the
analytical formula
\begin{equation}
    \ln \frac{\mathrm{N} _{col} (J)}{2 J + 1} = a \exp(-E(J)/b) + c + d\,E(J),
\end{equation}
where $E$ is the excitation energy of level $J$, and $a$, $b$, $c$ and $d$
are free parameters.  The population of the lowest $J=0,2,4,6,8$ levels
is represented by a linear fit corresponding to an excitation
temperature of 22.4 K.  Using this approach we have computed the total
C$_3$ column densities (see the last row of Table~\ref{TableNcol})
resulting in a value of N$_{col}= (6.61 \pm 0.19) \times 10^{12}$ cm$^{-2}$
for the weighted average over all bands.

The differences between the column densities determined for the
individual bands is representative for the overall quality of
our data and the procedures followed.  The total column density
concluded from the weighted average of all bands is shown in
Table~\ref{Table6} as a value representative for C$_3$ along with
column densities for other species.

From the present observations it is also possible to deduce a CH
column density toward HD~169454 from the lowest Q and R lines in the B-X
violet band at 3886.409 and 3890.217\,\AA, probing both lowest-$J$
$\Lambda$-doublet components, under the assumption that the lines are
optically thin.  The column density of CH$^+$ toward HD~169454 is obtained
from the 3597\,\AA\ transition in the A-X (1,0) band.  The resulting values
are included in Table~\ref{Table6}.

\begin{table}
\caption[]{Summary of observed molecular column densities towards HD~169454. }
% \label{TableSummary}
\label{Table6}
\begin{tabular}{@{}cccc}
\hline
Molecule & N$_{col}$             &  T$_{exc}$  & Source \\
         & ($10^{12}$ cm$^{-2}$) &  (K)        &  \\
\hline
C$_{2}$ &   65$\pm$1      &  19$\pm$2 & 1\\
        &   73$\pm$14     &  15$^{+10}_{-5}$ & 2 \\
        &   70$\pm$14     &   & 3 \\
        &  160$\pm$29     &   & 4  \\
C$_{3}$ & 6.61$\pm$0.19   &  22.4$\pm$1.0 & 5 \\
        & 2.24$\pm$0.66$^{a}$   & 23.4$\pm$1.4 & 4 \\
        &  4.5$\pm$0.3$^{b}$    &  & 3 \\
CH      & 39.6$\pm$0.3$^{c}$      &  & 5 \\
        &   46$\pm$8           &  & 2 \\
        &   36.5$^{+12.6}_{-7.8}$   &  & 6 \\
CH$^+$     &   20.8$\pm$0.2$^{e}$    &  & 5 \\
H$_{2}$ &   (8$\times$10$^{20}$$^{d}$)      &  & 5 \\
\hline
\end{tabular}
\\
$^a$ based on a sample of rotational lines with $J \leq 8$;\\
$^b$ based on unresolved rotational lines;\\
$^c$ based on CH line at 3886.409 \AA;\\
$^d$ based on correlation of CH w.r.t. H$_2$ (\citet{Weselak2004}; Fig. 3. therein);\\
$^e$ based on the CH$^+$ line at 3957\,\AA.\\
Source: (1)~\citet{Kazmierczak2010};
(2) \citet{Jannuzi1988};
(3) \citet{Oka2003};
(4) \citet{Adamkovics2003};
(5) This work;
(6) \citet{Crawford1997}
\end{table}

\section{Excitation models}

Recently, \citet{Roueff2002} (see also \cite{Welty2013}) presented an
excitation model of C$_3$ towards HD~210121.  An interesting aspect of
their approach was the inclusion of destruction and formation processes
of C$_3$, in view of its short life time in typical diffuse clouds.
As a consequence, the initial population of the highest rotational levels in
the formation process may not change significantly by collisions
before the molecule is destroyed by photodissociation.  For a quantitative
estimate of the population distribution two crude approximations were
made.  First, it was assumed that the destruction rate of C$_{3}$ is
the same in each energetic state.  Second, C$_{3}$ is assumed to be formed in
rotational states following a Boltzmann distribution characterized by a
formation temperature, $T_f$.

For the analysis of our data, we have performed computations strictly
following this approach outlined by~\citet{Roueff2002} using
their original collisional rates in the first set of models and modified
collisional rates as described below in a second set of models. The
calculations are performed with the RADEX code of~\citet{RADEX}.
Minor modifications are necessary to include production and
destruction terms into the statistical equilibrium equations.

There is some degeneracy in the model parameters providing good fits to the
observed populations.  A one parameter family of models may be constructed
as a function of the gas density, here assumed to be composed of molecular
hydrogen.  The density $n_{H_2}$ is varied in a range between 400 and 5000
cm$^{-3}$, since it cannot be determined unambiguously.  
A $\chi^2$-test of the goodness-of-fit was used for the determination
of the best fits among the family of models assuming a 1 percent significance level.
Values below 400 cm$^{-3}$ are ruled out by the fit.  The upper bound of 5000 cm$^{-3}$
is chosen on physical grounds as a value significantly exceeding the gas
density in the sight line to HD~169454 as concluded from models describing
C$_2$ excitation \citep{Kazmierczak2010,Casu2012}, $n_{H_2} = 350-500$
cm$^{-3}$.
The models are also characterized by the gas kinetic temperature
$T_k$, the formation temperature $T_f$,
and the destruction rate $D$ of C$_3$ per particle in s$^{-1}$.  The
best fits for fixed densities $n_{H_2}$ are described by the following
sets of ($n_{H_2}$, $T_{k}$, $D$) parameters: (400 cm$^{-3}$, 12 K,
0.5$\times$10$^{-9}$ s$^{-1}$), (500 cm$^{-3}$, 13 K, $0.6 \times 10^{-9}$
s$^{-1}$), (1000 cm$^{-3}$, 16 K, 1$\times$10$^{-9}$ s$^{-1}$), and (5000
cm$^{-3}$, 18 K, 5$\times$10$^{-9}$ s$^{-1}$).
The formation temperature $T_f$ is not well determined and may vary
between 150 and 750 K, with a highest probability at 300 K.
% variedis determined at 300$^{+50}_{-30}$ K.
We estimate uncertainties of 2 K in the gas temperature and up to 30 \% in the
destruction rate for each model.  For higher densities above
$n_{H_2}=1000$ cm$^{-3}$, one observes that the population distribution
depends on the ratio of the collisional rate to the destruction rate,
$n_{H_2} q_{ul}/ D$, with $q_{ul}$ the rate coefficient.  A
representative model for $n_{H_2}=1000$ cm$^{-3}$ is shown in
Fig.~\ref{figrotdiagram} with the red line.  The distribution of column
densities in the other models cited above are virtually the same and are not
shown.

Recently, \citet{Abdallah2008} computed collisional rates of C$_{3}$
colliding with helium for low rotational states $J=0-10$ and for a gas
temperature in the range of 5 to 15\,K.  In our model the rate coefficients
are scaled by 1.38 to represent collisions with H$_{2}$ instead of He
and used in a second tier of calculations based on these collision rates.
De-excitations from higher levels are fixed at values from the highest
available $J=10$ rotational level. The de-excitation collisional
rates of rotational levels of the ground vibrational state are an order of
magnitude larger than the $q_{ul} =2 \times 10^{-11}$
cm$^{3}$\,s$^{-1}$ assumed by \citet{Roueff2002}.  As a result, the models
with the new collisional rates require a much higher destruction rate to fit
the observed column densities well.  This is shown in the two
exemplary models characterized by set of ($n_{H_2}$, $T_{k}$, $D$)
parameters: (500 cm$^{-3}$, 18 K, $12 \times 10^{-9}$ s$^{-1}$) and (1000
cm$^{-3}$, 18 K, $24 \times 10^{-9}$ s$^{-1}$). Also in this set of models
the broad range of formation temperatures between 150 and 750 K 
with peak probability around 250 K is acceptable. In a large range of densities the gas
temperature is virtually constant and amounts to 18 K with a higher uncertainty of 3 K.
The representative model corresponding to $n_{H_2}$=500 cm$^{-3}$ is shown in
Fig.~\ref{figrotdiagram} as a blue line.  Even if the conversion of
collisional rates to collisions with H$_{2}$ is inaccurate, collisions with
only He (20 percent of the molecular hydrogen density) require an increase
of the destruction rate $D$ by over a factor two with respect to the original
model of \citet{Roueff2002}.

The presented models cannot determine uniquely the gas density nor the
destruction rate or the formation temperature
without additional assumptions made for these parameters.
The first set of excitation models requires that $n_{H_2}$ must be higher
than 2000 cm$^{-3}$ to produce a gas kinetic temperature $T_k$ in agreement with the
excitation temperature $T_{exc}$, here 22\,K, derived from the low-$J$ curve in
the Boltzmann plot. The second set of models does not allow to put
any such constraints on the density.

% Excitation models suggest that {\bf the density aorund  500 cm$^{-3}$
% that the gas kinetic temperature may be lower than
% the excitation temperature, here 22\,K, derived from the low-$J$ curve in
% the Boltzmann plot.
% {\bf This is rather clear in the first set of models where gas temperature
% must be not higher than 16 K. The second set of models suggests gas kinetic temperature of
% 18~K, which is less than excitation temperature of C$_3$ 22~K derived
% in this paper and marginally less than the gas temperature taken as average of
% C$_2$ and C$_3$ excitations. The additional model with the gas kinetic temperature of 20 K
% is shown in Fig.~\ref{figrotdiagram} to illustrate the level of divergence
% with the observations.}

If the destruction of C$_{3}$ is mainly caused by photodissociation then the
photodissociation rate in the radiative field defined by \citet{Draine1978}
amounts to $5 \times 10^{-9}$\,s$^{-1}$ \citep{Dishoeck1988}.  Assuming that
the center of the cloud for a total extinction A$_{V}$ of 2.8-3.4 (see
section 7) is penetrated by such a radiation field and that in the UV-range
extinction is twice as high as in the visible range, a destruction
rate of $0.5 \times 10^{-9}$ follows.  Hence, a significant enhancement of
the radiation field is necessary to explain the results of models with
collisional rates based on computations by~\citet{Abdallah2008}.  Another
possibility, not further discussed here, is that other C$_{3}$ destruction
processes are decisive. Clearly, more detailed research on
collisional rates is necessary before the physical conditions of the
molecular gas can be deduced in detail from the C$_3$ column
densities.

\section{Discussion}

Since the first identification of C$_3$ in diffuse clouds by
\citet{Maier2001} estimations of C$_3$ abundances are prone to large
uncertainties due to low signal-to-noise ratios of the astronomical
observations.  Available determinations of column densities for C$_{3}$
towards HD~169454 are summarized in Table~\ref{Table6}.
\citet{Adamkovics2003} were able to extract populations of only five
rotational levels up to $J=8$. An estimate of the column density was
made by \citet{Oka2003} on the basis of the contour shape of the pile of
Q-lines.  In the present high resolution analysis individual transitions
are resolved in the \~{A}$^{1}\Pi_{u}$-- \~{X}$^{1}\Sigma_{g}^{+}$ 000--000
band and complementary information from the vibronic bands is used,
yielding accurate column densities for C$_3$.

The chemical formation pathway of C$_{3}$ in diffuse clouds is closely
related to two other molecules commonly observed in the optical range, C$_{2}$
and CH (see e.g. \citet{Oka2003}).  The C$_{2}$ molecule provides also
information on the gas temperature, analogous to C$_{3}$.

The column density of C$_2$ in the sight line to HD~169454 was analyzed
previously.  \citet{Kazmierczak2010} derived N$_{col}$(C$_2$) $= 10^{13}$
cm$^{-2}$ and a gas temperature of 19 K.  Recently, \citet{Casu2012}
redefined the C$_2$ excitation model by~\citet{Dishoeck1982}, and based on
new collisional rates of~\citet{Najar2008} it was suggested that this
diffuse cloud must be composed of two spatial components.  From an analysis
of the observational C$_2$ data of~\citet{Kazmierczak2010}, they identified
a dense component with $n_{H_2} = 500$ cm$^{-3}$ and a more diffuse
component of 50 cm$^{-3}$ assuming background galactic radiation as
in~\citet{Dishoeck1982}.  The gas temperatures in the two components were
estimated at 20 and 100 K, respectively.  The parameters of the dense
component are very close to those determined by \citet{Kazmierczak2010},
$n_{tot} = 330$ cm$^{-3}$ and $T_{kin}$ = 19 K, using the original model by
\citet{Dishoeck1982}.

Column densities of CH based on the strong 4300\,\AA\ line in the sight line
to HD~169454 have been reported previously (e.g. \cite{Jannuzi1988}).
Here we determine a column density from the weaker CH B-X system, which is
in agreement with the previous result, but more accurate.

The column density of molecular hydrogen towards HD~169454 was never directly
determined.  We estimate N(H$_2$) at $8 \times 10^{20}$ cm$^{-2}$
indirectly from the correlation of H$_2$ with CH using a relation found by
\citet{Weselak2004}.
The total column density of N(C$_3$) = $6.6 \times 10^{12}$ cm$^{-2}$ then
corresponds to a fractional abundance
C$_3$/H$_2$ = $8.2 \times 10^{-9}$.

The present determination of the collisional temperature of $22 \pm 1$ K for C$_3$
compares well with the value of $19 \pm 2$ K obtained for C$_2$ towards
the same line of sight \citep{Kazmierczak2010}. Excitation models for C$_3$
\citep{Roueff2002} and for C$_2$ \citep{Dishoeck1982}
assume that the lowest rotational levels from which the excitation
temperature is determined are close to thermal.
This suggests that the gas temperature for both species should be about the same,
i.e. $21 \pm 2$~K. A detailed analysis in terms
of excitation models for C$_3$ shows that the value of 21 K lies in acceptable
range of parameters, with some exceptions for when the gas density $n_{H_2}$ is lower than 2000 cm$^{-3}$ in the first set of models. At lower gas densities of 350-500 cm$^{-3}$,
as suggested by the analysis of excitation of C$_2$ towards HD~169454 \citep{Kazmierczak2010,Casu2012}, a lower value for the gas temperature is necessary.
% suggests however a lower value for the gas temperature with an upper
% limit at $18 \pm 1$ K.
This may be attributed to the fact that C$_3$ should form in the
central part of the diffuse cloud toward HD~169454, at a slightly lower gas kinetic temperature.
In view of the many uncertainties in the prevailing excitation models
we consider these values as in acceptable agreement.

The derived value of the gas temperature is close to that found in detailed models
of diffuse molecular clouds including processes on grains~\citep{Hollenbach2009}.
For A$_V$ = 1.7 the gas kinetic temperature for
$n=10^3$ cm$^{-3}$ and radiation field expressed in units of the Draine field
intensity G$_0$ = 1 is in the range between 15 and 20 K (see
Fig. 11  in \citet{Hollenbach2009}) and similarly for $n=10^4$ cm$^{-3}$
and G$_0$=10 (see Fig. 7 in \citet{Hollenbach2009}).
The excitation model for C$_2$ allows an estimation
of the ratio of $n$ to G$_0$ and does not constrain the radiation field.

\citet{Crawford1997} observed C$_2$ and CH (as well as CN) towards HD~169454
detecting two slightly separated (by $0.7 \pm
0.1$ km s$^{-1}$) velocity components in the C$_2$ lines. The possible presence
of multiple clouds in the sight line does not impair the obtained results,
because the lines are optically thin and the excitation conditions  and gas
densities derived by \citet{Crawford1997} are quite similar. However, presence of
multiple clouds may influence the chemical analysis, effectively
lowering the optical depth to the external radiation field in each cloud.

\section{Interstellar cloud towards HD~169454}

The observational data offer another opportunity to analyze the chemical
abundances of small carbon chains in a molecular cloud in the sight line to
HD~169454.  Such a model was presented in an extensive analysis by
\citet{Jannuzi1988}.  Here we use the Meudon PDR code \citep{lePetit2006}
for the computation of the chemical composition of the plane-parallel cloud
irradiated by the average galactic background radiation field.

HD~169454 is a B1 Ia star \citep{Mendoza1958} with B-V determined at 0.90
\citep{Fernie1983}.  This translates to E(B-V) = 1.09 according to
\citet{Papaj1993}.  Assuming the typical total to selective extinction ratio
R=3.1, we find that A$_{V,tot}$=3.4.  The standard mean Galactic extinction
curve \citep{Fitzpatrick1990} is used in the model, which seems to
reproduce the extinction curve towards HD~169454 \citep{Wegner2002}
well.  We further assume, that the translucent cloud is irradiated from
both sides by the far ultraviolet radiation field.  The resulting
field in the central part of the cloud of total thickness A$_{V,tot}$ is
then reduced by extinction of only half of that. In the case of
asymmetric illumination, or of a special arrangement of the cloud in the
line of sight, A$_V$ may be even smaller.

\begin{figure}
\centering
\includegraphics[width=8.4cm]{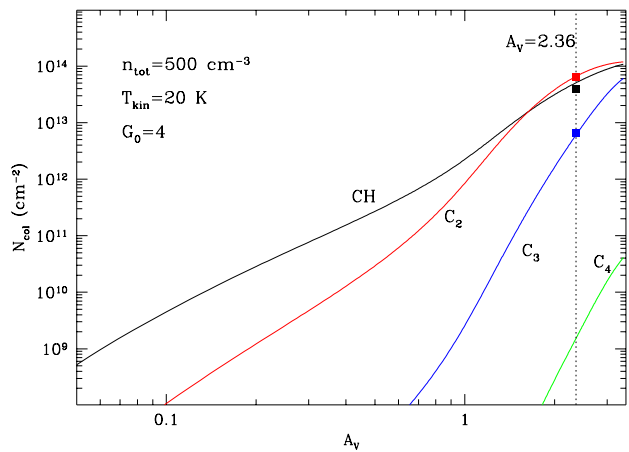}
\caption{Abundances of CH, C$_2$, C$_3$, and C$_4$ carbon species
computed as a function of the total visual absorption A$_V$.
}
\label{FigChem}
\end{figure}

We have performed a calculation for a single cloud with a total density of 500 cm$^{-3}$ and a gas
temperature of 20~K illuminated by the radiation field enhanced with
a factor G$_0$ of 4 relative to the standard value of
\citet{Draine1978}. The results are shown in Fig.~\ref{FigChem}.
With these parameters we reproduce the observed abundance
ratio of C$_2$/C$_3$ $\sim$ 10  for A$_V$=2.36.
The prediction for the abundance of C$_4$
at this A$_V$ value is two orders lower than the abundance of C$_3$
suggesting that more observational efforts are necessary to detect
also C$_4$ in
translucent clouds.  The calculated abundance of CH is slightly higher
than the observed value. Assuming that the optical extinction
A$_V$ lower than the value is estimated from E(B-V), good agreement
may be reached also for other combinations of A$_V$ and G$_0$: e.g.
A$_V$=1.86 and G$_0$=2, A$_V$=2.66 and G$_0$=6.

It is a very complex task to make reasonable assumptions
concerning the internal structure of any interstellar cloud.
Clouds are likely non-homogeneous and asymmetrically irradiated by
neighbouring stars. Another problem is the dust content.
As demonstrated recently by \citet{Krelowski2012}
the abundance ratio of the simple radicals CH
and CN also depends on the shape of the extinction curve.

\begin{figure}
\centering
\includegraphics[width=8.4cm]{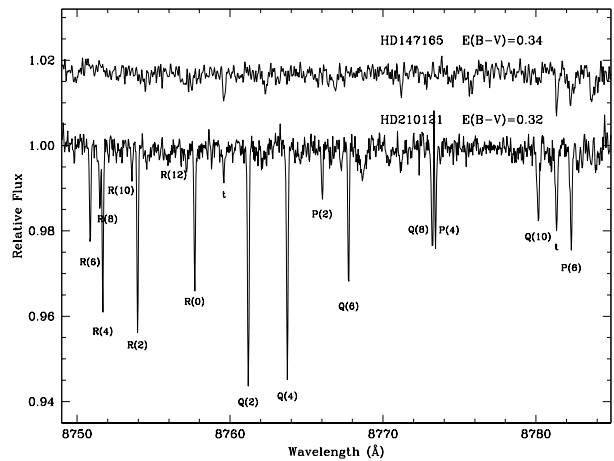}
\caption{Spectral range of C$_{2}$ Phillips (2,0) band towards two
objects HD~210121 and HD~147165 with similar colour excesses
E(B-V) and different abundances of C$_2$. }
\label{FigC2}
\end{figure}

To illustrate the strong dependence on far-UV extinction we display spectra
with very distinct C$_2$ spectral features toward two stars that have
practically the same E(B-V).  The data, shown in Fig.~\ref{FigC2}, are
obtained from our present observation (HD~147165 in ESO Program
082.C-0566(A)) and from the ESO Archive (HD~210121 in Program 71.C-0513(C)).
These spectra show that the C$_2$ abundance is drastically lower for the
diffuse cloud toward HD~147165, which undergoes a low far--UV extinction
field, than for the cloud towards HD~210121, which undergoes a high far--UV
extinction field.  The far-UV extinction curves can be found in Figs.  4.20
and 4.31 in the survey by \citet{Fitzpatrick2007}.

\section{Conclusion}

We report on a high-quality absorption spectrum, in terms of both resolution
and signal-to-noise ratio, of the C$_3$ molecule towards HD~169454.  Besides
a fully resolved spectrum of the \~{A} -- \~{X} 000 - 000 band at $\sim
4052$~\AA\ seven further vibronic C$_3$ bands were identified in the
range 3793-4000 \AA.  These absorptions had been suspected previously
\citep{Gausset1965}, but now these bands are unambiguously detected
for the first time with UVES-VLT.  Four of those vibronic bands have been
observed as well along the sight lines to two heavily reddened stars:
HD~154368 and HD~73882.  The observations are supported by laboratory
measurements of all eight bands under high resolution using cavity
ring-down laser spectroscopy.

The accurate value for the C$_3$ column density in the translucent cloud
towards HD~169454 was included in an excitation model, applying the Meudon
PDR-code by \citet{Roueff2002}, yielding good agreement for the column
densities of rotational levels.  The calculations do not allow to
determine gas density and destruction rate uniquely.  As far as the rate
coefficients and photodissociation rates are established, the model depends
on the ratio of gas density to intensity of radiation field.  When modifying
the model with recently updated collisional rates~\citep{Abdallah2008} the
model results in high destruction rates required for C$_{3}$, inconsistent
with the present understanding of the destruction process.

Observation of carbon chain molecules in optical spectra of diffuse
clouds was up to now unsuccessful, except for C$_3$.  All complex organic
molecules identified in the interstellar space have been found in dense
protostellar clouds.  The detection and identification of a number of very
weak absorptions of the C$_3$ molecule raises hope that the forest of narrow
absorptions in the blue-violet part of the spectrum towards reddened stars
may uncover assignments of heavier carbon species.  Many weak as yet
unidentified spectral features appear to be molecular lines, rather than
noise.  The regular pattern of C$_3$ bands allows for an unambiguous
identification, as shown in the present study.  Heavier species like C$_4$
and C$_5$ would produce more compact patterns due to the smaller
rotational constant or due to a strong Q-branch compared to R-
and P-branches characteristic for $\Sigma$--$\Pi$ transitions in case of
C$_5$, particularly for low excitation temperature, which may lead to their
detection.  The assignment of absorption features in the UV-blue spectral
range to the C$_3$ molecule is of relevance in the context of searches for
carriers of the diffuse interstellar bands (DIBs).  While most of the
absorption features detected in translucent clouds in sight-lines toward
reddened stars are ascribed to DIBs of unknown origin, the presently
observed features can be excluded from the DIB-lists for which carriers are
sought.

\section*{acknowledgements}
We are indebted to J.~H.~Black for providing us
with detailed information on the collisional model for C$_{3}$.  
J. Kre{\l}owski acknowledges financial support from the Polish National Center
for Science during the period 2011 - 2013 (grant UMO-2011/01/B/ST2/05399).  
G.A. Galazutdinov acknowledges support of Chilean fund FONDECYT-regular
(project 1120190).  M.~R.~Schmidt acknowledges support by the National
Science Center under grant (DEC-2011/01/B/ST9/02229).  W. Ubachs
acknowledges support from the Templeton Foundation.  The authors situated in
the Netherlands acknowledge support through FOM, NOVA, and the Dutch
Astrochemistry Program.  We are grateful for the assistance of the Paranal
Observatory staff members.
The data analysed here are based on
observations made with ESO Telescopes at the Paranal
Observatory under programmes 71.C-0367(A), 076.C-0431(B) and
082.C-0566(A).

\label{lastpage}

\end{document}